%% file: QC_SE_PD.tex
\documentclass[journal]{./package_style/IEEEtran}
\usepackage[latin9]{inputenc}
\usepackage{draftwatermark}
\usepackage{bm}
\usepackage{soul}
\usepackage[dvipsnames]{xcolor}
\usepackage{amsmath}
\usepackage{amssymb}
\usepackage{graphicx}
\usepackage[percent]{overpic}
\usepackage{algorithmic}
\usepackage{array}
\usepackage{rotating}
\usepackage{bm}
\usepackage{lipsum}
\usepackage[percent]{overpic}
\usepackage{color}
\usepackage{pgfplots}
\pgfplotsset{compat=newest}
\pgfplotsset{plot coordinates/math parser=false}
\newlength\figureheight
\newlength\figurewidth 
\usepackage{amsfonts}
\usepackage{epstopdf}
\usepackage{cite}
\usepackage{cases}
\usepackage{anyfontsize}
\usepackage{mathtools}

\usepackage{multirow}%
\usepackage{hhline}%
\usepackage{xfrac}
\usepackage{mathtools}
\makeatletter
\let\MYcaption\@makecaption
\makeatother
\usepackage[font=footnotesize]{subcaption}
\makeatletter
\let\@makecaption\MYcaption
\makeatother

\SetWatermarkText{Preprint}
\begin{document}
\title{\huge Automatic Quality Control and Enhancement for Voice-Based Remote Parkinson's Disease Detection}

\author{{Amir Hossein Poorjam, \emph{Student Member, IEEE}, Mathew Shaji Kavalekalam, \emph{Student Member, IEEE}, Liming Shi, \emph{Student Member, IEEE}, Yordan P. Raykov, Jesper Rindom Jensen, \emph{Member, IEEE}, Max A. Little, \emph{Member, IEEE} and Mads Gr\ae sb\o ll Christensen, \emph{Senior Member, IEEE}} 
\thanks{This work was funded by Independent Research Fund Denmark: DFF 4184-00056.}
\thanks{A.H. Poorjam, M.S. Kavalekalam, L. Shi, J.R. Jensen and M.G. Christensen are with the Audio Analysis Lab, CREATE, Aalborg University, Aalborg 9000, Denmark (e-mail: \tt \{ahp,msk,ls,jrj,mgc\}@create.aau.dk).}

\thanks{M.A. Little is with the School of Engineering and Applied Science, Aston University, Birmingham, UK, and also with Media Lab, MIT, Cambridge, Massachusetts, USA (e-mail: \tt max.little@aston.ac.uk).}

\thanks{Y.P. Raykov is with the School of Engineering and Applied Science, Aston University, Birmingham, UK. (e-mail: \tt yordan.raykov@gmail.com).}
}
\maketitle

\begin{abstract}
The performance of voice-based Parkinson's disease (PD) detection systems degrades when there is an acoustic mismatch between training and operating conditions caused mainly by degradation in test signals. 
In this paper, we address this mismatch by considering three types of degradation commonly encountered in remote voice analysis, namely background noise, reverberation and nonlinear distortion, and investigate how these degradations influence the performance of a PD detection system.
Given that the specific degradation is known, we explore the effectiveness of a variety of enhancement algorithms in compensating this mismatch and improving the PD detection accuracy.
Then, we propose two approaches to automatically control the quality of recordings by identifying the presence and type of short-term and long-term degradations and protocol violations in voice signals.
Finally, we experiment with using the proposed quality control methods to inform the choice of enhancement algorithm.
Experimental results using the voice recordings of the mPower mobile PD data set under different degradation conditions show the effectiveness of the quality control approaches in selecting an appropriate enhancement method and, consequently, in improving the PD detection accuracy.
This study is a step towards the development of a remote PD detection system capable of operating in unseen acoustic environments.

\end{abstract}


\IEEEpeerreviewmaketitle

\section{Introduction}
\label{sec:intro}
Parkinson's disease (PD) is a neurodegenerative disorder which progressively makes the patients unable to control their movement normally and, consequently, decreases the patients' quality of life \cite{Ishihara2007}.
Since there is no cure for PD, it is necessary to develop tools to diagnose this disease in early stages in order to control its symptoms. 
Speech is known to reflect the PD symptoms since the majority of PD patients suffer from some forms of vocal disorder \cite{Ho1998}.
It has been demonstrated in \cite{Eliasova} that early changes of clinical symptoms of PD are more reflected and pronounced in acoustic analysis of voice signals than in perceptual evaluation of voice by a therapist. This has motivated researchers to take advantage of advanced speech signal processing and machine learning algorithms to develop highly accurate and data-driven methods for detecting PD symptoms from voice signals \cite{Tsanas2012,Zhan2016,Gil2006}.
Moreover, advances in smart phone technology provide new opportunities for remote monitoring of PD symptoms by bypassing the logistical and practical limitations of recording voice samples in controlled experimental conditions in clinics \cite{Rusz2018,Zhan2016}.
However, there is a higher risk outside controlled lab conditions that participants may not adhere to the test protocols, which probe for specific symptoms, due to lack of training, misinterpretation of the test protocol or negligence.
Moreover, voice signals in remote voice analysis might be subject to a variety of degradations during recording or transmission. Processing the degraded recordings or those which do not comply with the assumptions of the test protocol can produce misleading, non-replicable and non-reproducible results \cite{Fan2014} that could have significant ramifications for the patients' health. 
In addition, degradation of voice signals produces an acoustic mismatch between the training and operating conditions in automatic PD detection.
A variety of techniques have been developed for compensating this type of mismatch in different speech-based applications  \cite{Gong1995,Fakhry2018,Hansen2014,Alam2017,Mammone1996a,Nercessian2016,Poorjam2016} which can, in general, be categorized into four classes: (1) searching for robust features which parameterize speech regardless of degradations; (2) transforming a degraded signal to the acoustic condition of the training data using a signal enhancement algorithm\footnote{In this paper, by ``signal enhancement'', we refer to all algorithms intended to enhance the quality of degraded signals.}; (3) compensating the effects of degradation in the feature space by applying feature enhancement; and (4) transforming the parameters of the developed model to match the acoustic conditions of the degraded signal at operating time. 
However, to the best of the authors' knowledge, there is a lack of studies of the impact of acoustic mismatch and the effect of compensation on the performance of PD detection systems.
Vasquez-Correa et al. proposed a pre-processing scheme by applying a generalized subspace speech enhancement technique 
to the voiced and unvoiced segments of a speech signal to address the PD detection in non-controlled noise conditions \cite{Vasquez}. 
They showed that applying speech enhancement to the unvoiced segments leads to an improvement in detection accuracy while the enhancement of voiced segments degrades the performance. However, this study is limited in terms of degradation types as it only considered the additive noise. 
Moreover, they only evaluated the impact of an unsupervised enhancement method on PD detection performance, while the supervised algorithms have, in general, shown to reconstruct higher quality signals as they incorporate more prior information about the speech and noise.

Another open question which, to the authors' knowledge, has not been addressed is whether applying ``appropriate'' signal enhancement algorithms to the degraded signals will result in an improvement in PD detection performance.
Answering this question, however, requires prior knowledge about the presence and type of degradation in voice signals, which can be achieved by controlling the quality of recordings prior to analysis.
Quality control of the voice recordings is typically performed manually by human experts which is a very costly and time consuming task, and is often infeasible in online applications.
In \cite{Poorjam2019}, the problem of quality control in remote speech data collection 
has been approached by identifying the potential outliers which are inconsistent, in terms of the quality and the context, with the majority of speech samples in a data set. Even though very effective in finding outliers, it is not capable of detecting the type of degradation nor identifying short-term protocol violations in recordings.
To identify the type of degradation in pathological voices, Poorjam et al. proposed two different parametric and non-parametric approaches to classify degradations commonly encountered in remote pathological voice analysis into four major types, namely background noise, reverberation, clipping and coding \cite{Poorjam2017,poorjam2018parametric}. 
However, the performance of these approaches is limited when new degradation types are introduced. 
Furthermore, the presence of outlier recordings, which do not contain relevant information for PD detection due to long-term protocol violations, is not considered in these methods and, therefore, there is no control over the class assignment for such recordings.
To address the frame-level quality control in pathological voices, Badawy et al. proposed a general framework for detecting short-term protocol violations using a nonparametric switching autoregressive model \cite{Badawy2018}. In \cite{Poorjam2019a}, a highly accurate approach for identifying short-term protocol violations in PD voice recordings has been proposed which fits an infinite hidden Markov model to the frames of the voice signals in the mel-frequency cepstral domain. However, these two approaches do not identify short-term degradations (e.g. the presence of an instantaneous background noise) in voice signals.

To overcome the explained limitations in the existing methods, we propose two approaches for controlling the quality of pathological voices at recording-level and frame-level in this paper. 
In the recording-level approach, separate statistical models are fitted to the clean voice signals and the signals corrupted by different degradation types. 
The likelihood of a new observation given each of the models is then used to determine its degree of adherence to each class of acoustic conditions.
This gives us the flexibility not only to associate multiple classes to a voice signal corrupted by a combination of different degradations, but also to consider a recording as an outlier or a new degradation when it is rejected by all the models.
In the frame-level approach, on the other hand, we extend the work in \cite{Poorjam2019a} to identify short-term protocol violations and degradations in voice signals at the same time.
We show how the proposed quality control approaches can effectively inform the choice of signal enhancement methods 
and, consequently, improve the PD detection performance.
The contribution of this paper is thus three-fold: (1) we investigate the impact of acoustic mismatch between training and operating conditions, due to degradation in test signals, on the PD detection performance; (2) to identify this mismatch, we propose two different approaches to automatically control the quality of pathological voices at frame- and recording-level; and (3) to efficiently reduce this mismatch, given that the specific degradation is known, we explore a variety of state-of-the-art enhancement algorithms and their effectiveness in improving the performance of a PD detection system.
The rest of the paper is organized as follows. Section \ref{sec:pd_detection} explains the PD detection system that we have used for the experiments throughout this paper. In Section \ref{sec:degradation_impact}, we investigate the impact of three major types of signal degradation commonly encountered in remote voice analysis, namely noise, reverberation and nonlinear distortion, on the performance of the PD detection system. Following that, in Section \ref{sec: enhancement}, we investigate on the influence of noise reduction and dereverberation algorithms on the performance of the PD detection system. 
In Section \ref{sec: degradation_det}, we propose two different quality control approaches and investigate how these methods can improve the performance of PD detection. Finally, Section \ref{sec:conclusion} summarizes the paper.
\section{Parkinson's Disease Detection System}
\label{sec:pd_detection}
In this section, we describe the PD detection system we will use for further quality control and enhancement experiments.
This approach, which was proposed in \cite{Moro-Velazquez2018}, fits Gaussian mixture models (GMMs) to the frames of the voice recordings of the PD patients and the healthy controls (HC) parametrized by perceptual linear predictive (PLP) coefficients \cite{Hermansky1990}.
The motivation for using PLP parametrization is that the perceptual features are more discriminative in PD detection than the conventional and clinically interpretable ones (such as standard deviation of fundamental frequency, jitter, shimmer, harmonic-to-noise ratio, glottal-to-noise exitation ratio, articulation rate, and frequencies of formants), particularly when the voice is more noisy, aperiodic, irregular and chaotic which typically happens in more advanced stages of PD \cite{Orozco-Arroyave2013,Brabenec2017,Mekyska2016}.

Acoustic features of the PD patients' recordings and those of the healthy controls are modeled by GMMs with the likelihood function defined as:
\begin{equation}\label{eq:gmm_lh_function}
    p(\bm{x}_t|\lambda)=\sum_{c=1}^{C}b_c p(\bm{x}_t|\bm{\mu}_c,\bm{\Sigma}_c),
\end{equation}
where $\bm{x}_t$ is the feature vector at time frame $t$, $b_c$ is the mixture weight of the $c^{\mathrm{th}}$ mixture component, $C$ is the number of Gaussian mixtures, $p(\bm{x}_t|\bm{\mu}_c,\bm{\Sigma}_c)$ is a Gaussian probability density function where $\bm{\mu}_c$ and $\bm{\Sigma}_c$ are the mean and covariance of the $c^{\mathrm{th}}$ mixture component, respectively. The parameters of the model, $\lambda=\{b_c,\bm{\mu}_c,\bm{\Sigma}_c\}_{c=1}^C$, are trained through the expectation-maximization algorithm \cite{Reynolds1995}. 

Given $\bm{X}=(\bm{x}_1,\ldots,\bm{x}_t,\ldots,\bm{x}_T)$, a sequence of feature vectors, the goal in PD detection is to find the model which maximizes $p(\lambda_j|\bm{X})$, where ${j\in\lbrace\mathrm{PD},\mathrm{HC}\rbrace}$. Using the Bayes' rule, independence assumption between frames, and assuming equal priors for the classes, the PD detection system computes the log-likelihood ratio for an observation as:
\begin{equation}\label{eq:llk_pd}
    \sigma(\bm{X})=\sum_{t=1}^T\log p(\bm{x}_t|\lambda_{\mathrm{PD}})-\sum_{t=1}^T\log p(\bm{x}_t|\lambda_{\mathrm{HC}}).
\end{equation}
The final decision about the class assignment for an observation is made by setting a threshold over the obtained score.


\subsubsection{Experimental Setup}\label{ssec: pd_setup}
In this study, we use the sustained vowel /a/ as the speech material for PD detection since they provide a simpler acoustic structure to characterize the glottal source and resonant structure of the vocal tract than running speech. Moreover, perceptual analysis of different vowels suggests that the best PD detection performance can be achieved when the sustained vowel phonation /a/ is parametrized by the PLP features \cite{Orozco-Arroyave2013}.
We consider the mPower mobile Parkinson's disease (MMPD) data set \cite{Bot2016} which consists of more than 65,000 samples of 10 second sustained vowel /a/ phonations recorded via smartphones by PD patients and healthy speakers of both genders from the US.
The designed voice test protocol for this data set required the participants to hold the phone in a similar position to making a phone call, take a deep breath and utter a sustained vowel /a/ at a comfortable pitch and intensity for 10 seconds.
A subset of 800 good-quality voice samples (400 PD patients and 400 healthy controls equally from both genders) have been selected from this data set.
It should be noted that the health status in this data set is self-reported. To have more reliable samples, among participants who self-reported to have PD, we selected those who claimed that they have been diagnosed by a medical professional with PD and recorded their voice right before taking PD medications. For the healthy controls, we selected participants who self-reported being healthy, do not take PD medications, and claimed that they have not been diagnosed by a medical professional with PD.
All speakers of this subset had an age range of 58 to 72.
The mean $\pm$ standard deviation (STD) of the age of PD patients and healthy controls are 64$\pm$4 and 66$\pm$4, respectively. 
For all experiments in this paper, we downsampled the recordings from 44.1 kHz to 8 kHz since the enhancement algorithms used in this work are operating at 8 kHz.
To extract the PLP features, voice signals are first segmented into frames of 30 ms with 10 ms overlap using a Hamming window.
Then, 13 PLP coefficients are computed for each frame of a signal.
To consider the dynamic changes between frames due to the deviations in articulation, a first- and a second-order orthogonal polynomials are fitted to the two feature vectors to the left and right of the current frame. These features, which are referred to as \emph{delta} and \emph{double-delta}, were appended to the feature vector to form a 39-dimensional vector per each frame.
The number of mixture components for the GMMs were set to 32.
\subsubsection{Results}\label{ssec: pd_results}
\begin{figure}[tb]
\centerline{\includegraphics[width=0.35\textwidth, trim = 51mm 146mm 52mm 47mm, clip]{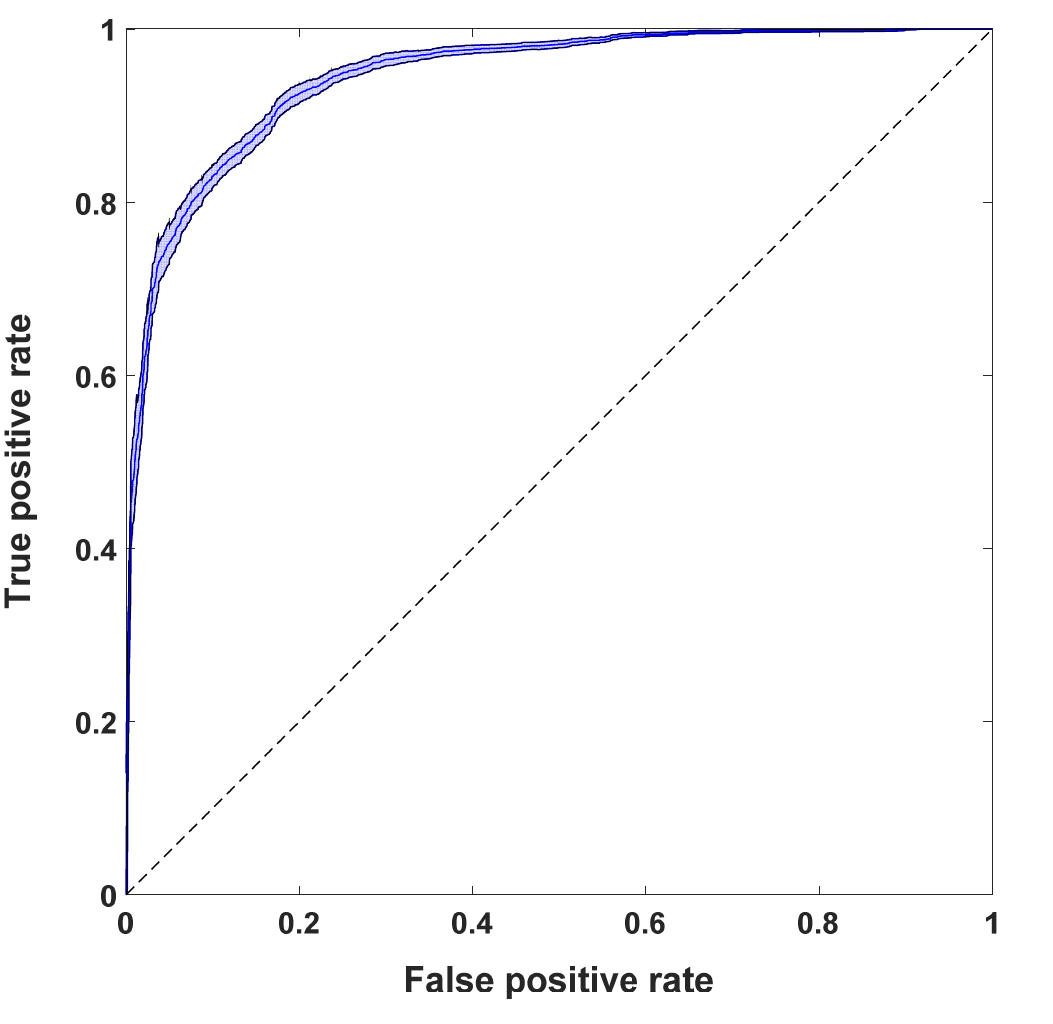}}
\caption{The ROC curve of the PD detection system, along with 95\% confidence interval shaded in blue. The dashed line shows the chance level.}
\label{fig:pd_clean}
\end{figure}

To evaluate the performance of the PD detection system in a matched acoustic condition, we used 5-fold cross validation (CV) in which the recordings were randomly divided into 5 non-overlapping and equal sized subsets. 
The entire CV procedure was repeated 10 times to obtain the distribution of detection performance. 
Fig.~\ref{fig:pd_clean} shows the performance in terms of the receiver operating characteristic (ROC) curve, along with 95\% confidence interval. 
In an ROC curve, the true positive rate is plotted against the false positive rate for different decision thresholds. 
The area under the curve (AUC) summarizes the ROC curve and represents the performance of a detection system by a single number between 0 and 1; the higher the performance, the closer the AUC value is to 1. Comparing with the commonly used classification accuracy, the AUC is a more preferred metric in this paper since it is a summary of the class overlap which sets a fundamental limit to the classification accuracy.
The mean AUC for this PD detection system is 0.95.

\section{Impact of Signal Degradation on PD Detection}
\label{sec:degradation_impact}
The PD detection system explained in the previous section gave a mean AUC of 0.95 in a matched acoustic condition. That is, when it was trained and tested using the clean recordings.
However, as alluded to in the introduction, recordings collected remotely in an unsupervised manner are seldom clean as they are often degraded by different types of degradation.
In this section we investigate the effect of 3 different commonly encountered degradations, namely  reverberation, background noise and nonlinear distortion on the performance of the PD detection system. 
It should be noted that even though we tried to choose the most reliable samples from the MMPD data set, the labels might still not be 100\% reliable as the diagnosis is self-reported. For this reason, we are more interested in how the relative PD detection performance is influenced systematically under application of different experimental conditions.

\subsection{Reverberation}
\label{ssec: reverberation_impact}
Reverberation is a phenomenon that occurs when the signal of interest is captured in an acoustically enclosed space. 
Apart from the direct component, the microphone receives multiple delayed and attenuated versions of the signal, which is characterized by the room impulse response (RIR). 
A metric commonly used to measure the reverberation is the reverberation time (RT60) \cite{vorlander2007auralization}. 
The presence of reverberation 
has shown to degrade the performance of speech-based applications such as speech and speaker recognition \cite{yoshioka2012making,Castellano1996}.
In this section, we investigate the effect of reverberation on the PD detection performance. 
To this aim, we used 5-fold CV repeated 10 times to evaluate the performance. In each iteration, the model was trained using the clean recordings of the training subset, and evaluated on the recordings of the disjoint test subset which were filtered with synthetic room impulse responses of RT60 varying from 300 ms to 1.8 s in 300 ms steps measured at a fixed position in a room of dimension 10 m $\times$ 6 m $\times$ 4 m. 
The distance between source and microphone is set to 2m. 
The room impulse responses were generated using the image method \cite{Allen1979} and implemented using the RIR Generator toolbox \cite{habets2006room}. 
Fig.~\ref{fig: auc_exp1_RT60} shows the impact of reverberation on the PD detection performance in terms of the mean AUC along with 95\% confidence intervals. We can observe from the plot that the PD detection system exhibits lower performance in reverberant environments, as expected, and the amount of degradation is related to the RT60.

\input{figures/degradations.tex}

\subsection{Background Noise}
\label{ssec: noise_impact}
Background noise is one of the most common types of degradation occurring during remote voice analysis. In this section we restrict ourselves to additive background noise and investigate how this can influence the PD detection performance.
To this aim, we performed the same CV procedure used for evaluating the impact of reverberation (explained in the previous section). In each iteration, the model was trained using the clean recordings of the training subset, and evaluated on the recordings of the test subset contaminated by an additive noise. The entire procedure was repeated for four different noise types, namely babble, restaurant, office and street noise\footnote{The babble, restaurant and street noise files have  been taken from https://www.soundjay.com/index.html and the office noise has been taken from https://freesound.org/people/DavidFrbr/sounds/327497} and different signal-to-noise ratios (SNRs) ranging from -5 dB to 10 dB in 5 dB steps.
Fig.~\ref{fig: auc_exp1_SNR} illustrates the impact of different noise types and different SNR conditions on the performance of the PD detection system in terms of the mean of AUC along with the 95\% confidence intervals.
We can observe a similar trends for all noise types that that the PD detection performance decreases as the noise level increases.

\subsection{Clipping}
\label{ssec: clipping}
In remote voice analysis, nonlinear distortion can manifest itself in speech signals in many different ways such as clipping, compression, packet loss and combinations thereof.
Here, we consider clipping as an example of nonlinear distortion in signals which is caused when a signal fed as an input to a recording device exceeds the dynamic range of the device \cite{eaton2013detection}. 
By defining the clipping level as a proportion of the unclipped peak absolute signal amplitude to which samples greater than this threshold are limited, we can investigate the impact of clipping on the PD detection performance. To this aim, the clean recordings of the test subset in each iteration of the CV were clipped with different clipping levels ranging from 0.1 to 0.8 in 0.1 steps.
Fig.~\ref{fig: auc_exp1_CLIP} shows the performance as a function of clipping level. Similar to the other types of degradation, it can be observed that increasing the distortion level in voice signals decreases the PD detection performance.

\section{Impact of Noise Reduction and Dereverberation on PD Detection}
\label{sec: enhancement}
As seen in Section \ref{sec:degradation_impact}, the degradation introduced to the signals can lead to reduction in the performance of the  PD detection system. 
Since there are practically an infinite number of possible types and combinations of nonlinear distortion that can be present in a signal, and since there is a lack of well-documented algorithms for dealing with most of the distortions (even in isolation), in this section, we only consider the degradations for which there are well-documented and verified enhancement algorithms such as noise reduction and dereverberation and investigate the effects of these algorithms on the PD detection performance.
To this end, from the 50 PD detection models developed and evaluated through 10 iterations of the 5-fold cross-validation procedure, as explained in Section (\ref{ssec: pd_results}), we selected one of the two models which showed the median performance and used it for further enhancement experiments in this section.
We have used a total of 160 recordings for testing the algorithms used in this section.
We will restrict ourselves to single channel enhancement algorithms. 
It should be noted that there exist a variety of objective and subjective metrics to measure the quality of the enhanced speech signal such as SNR, signal-to-distortion ratio \cite{Vincent2007}, perceptual evaluation of speech quality \cite{hu2007subjective} and short-time objective intelligibility \cite{Taal2010}. However, since our main goal in this work is to study the influence of speech enhancement on the PD detection performance, we evaluate the effectiveness of the algorithms in terms of the AUC.

\subsection{Dereverberation}
\label{ssec: dereverb}

Some of the  popular classes of dereverberation techniques are the spectral enhancement methods \cite{Habets2007}, probabilistic model based methods \cite{jukic2014speech,jukic2015multi} and inverse filtering based methods \cite{kameoka2009robust, huang2003class}. 
Spectral enhancement methods  estimate the clean speech spectrogram by frequency domain filtering using the estimated late reverberation statistics. 
The probabilistic model based methods model the reverberation using an autoregressive (AR) process, and the clean speech spectral coefficients using a certain probability distribution function. 
The estimated parameters of the model are then used to perform dereverberation. 
Lastly, the inverse filtering methods use a blindly estimated room impulse response to design an equalization system. 
These methods, which are mainly developed for the running speech, assume that the signal at a particular time-frequency bin is uncorrelated with the signals at that same frequency bin for frames beyond a certain number \cite{jukic2015multi}. However, this assumption is not valid for the sustained vowels which makes the dereverberation of the sustained vowels more challenging.
Recently, deep neural network (DNN) based dereverberation algorithms have gained attention \cite{han2015learning, santos2018speech} since they relax the assumption of uncorrelated neighboring time-frequency bins. The underlying principle of the DNN-based methods is to train a DNN to map the log-magnitude spectrum of the degraded speech to that of the desired speech. 

\begin{figure}[tb]
\centering
  \includegraphics[width=0.42\textwidth]{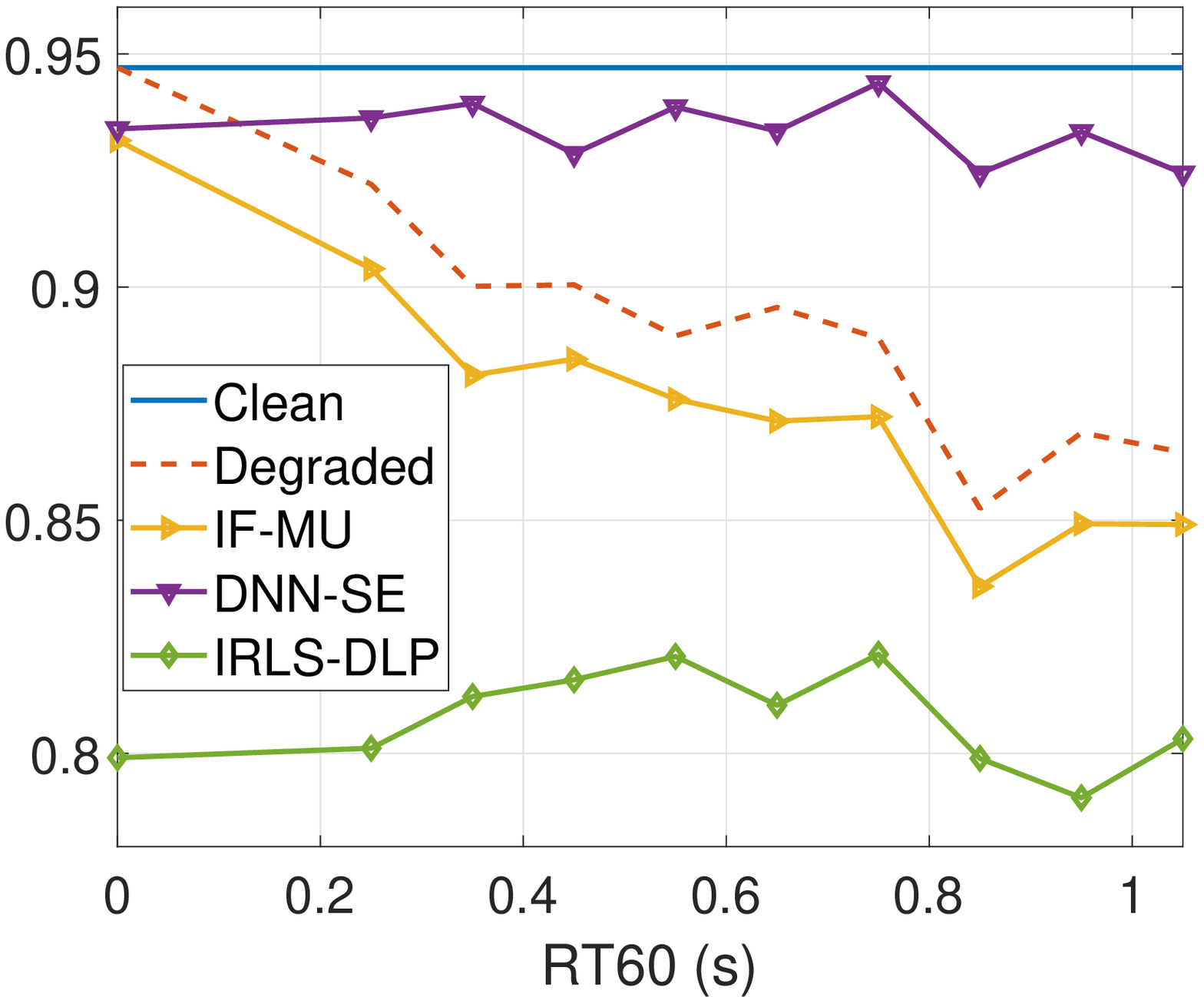}
  \vspace{-1mm}
  \caption{Impact of different dereverberation algorithms on the PD detection performance, in terms of AUC}
  \label{fig: auc_dereverb}
\vspace{-1mm}
\end{figure}

In this section, we investigate the effectiveness of different dereverberation algorithms in improving the PD detection performance.  
For dereverberation experiments, we used three different algorithms: a probabilistic model based algorithm proposed in \cite{jukic2015multi} (denoted as WPE-CGG, weighted prediction error with complex generalized Gaussian prior), an algorithm based on the inverse filtering of the modulation transfer function \cite{kameoka2009robust} (denoted as IF-MU, inverse filtering with multiplicative update), and a DNN-based speech enhancement algorithm proposed in \cite{han2015learning} (denoted as DNN-SE). It should be noted that the WPE-CGG and the IF-MU are unsupervised methods whereas DNN-SE is a supervised method. 
For the DNN-based algorithm, a feedforward neural network with 3 hidden layers of 1,600 neurons was used. To take into account the temporal dynamics, features of 11 consecutive frames (including the current frame, 5 frames to the left and 5 frames to the right over time) were provided to represent the input features of the current frames.
To train the DNN model, we selected 640 clean recordings from the MMPD data set and filtered them with the synthetic room impulse responses of RT60 ranging from 200 ms to 1 s in steps of 100 ms using the implementation in \cite{habets2006room} for a particular source and receiver position in a room of dimensions 10~m $\times$ 6 m $\times$ 4 m. 
For testing, the position of the receiver was fixed while the position of the source was varied randomly from 60 degrees left of the receiver to 60 degrees right of the receiver. 
Fig.~\ref{fig: auc_dereverb} shows the performance of the PD detection in terms of AUC for the different dereverberation algorithms. 
It can be observed from the figure that only DNN-SE is able to improve the PD detection performance while the other two methods degrade the performance. 
This is mainly due to two reasons: first, the DNN-SE is a supervised algorithm while the WPE-CGG and IF-MU are unsupervised; and second, the underlying assumption of the two unsupervised algorithms does not hold for the sustained vowels.
We have also included the case of zero RT60 to investigate the impact of processing of the clean recordings by these dereverberation algorithms.

\subsection{Noise reduction} \label{ssec: noise_reduction}

Methods for performing noise reduction can be broadly categorized into supervised and unsupervised methods. 
Unsupervised methods do not assume any prior knowledge about identity of the speaker or noise environment. 
The supervised methods, on the other hand, make use of training data to train the models representing the signals of interest or the noise environment. 
Some of the popular classes of supervised speech enhancement methods include the codebook-based methods \cite{srinivasan2006codebook, he2017multiplicative}, non-negative matrix factorization based methods \cite{mohammadiha2013supervised,Fakhry2018} and the DNN-based methods \cite{wang2018supervised}. In the supervised method, the speech and noise statistics/parameters estimated using the training data are exploited within a filter to remove the noise components from the noisy observation.
In this section, we used two supervised methods and one unsupervised method to investigate the effect of different noise reduction algorithms in reducing the acoustic mismatch between training and operating conditions.

The first supervised enhancement algorithm is based on the framework proposed in \cite{kavalekalam2019model}.
In this approach, a Kalman filter, which takes into account the voiced and unvoiced parts of speech \cite{goh1999kalman}, is used for enhancement. The filter parameters consist of the AR coefficients and excitation variance corresponding to speech and noise along with the pitch parameters (i.e. the fundamental frequency and the degree of voicing).
Based on \cite{kavalekalam2019model}, the AR coefficients and excitation variance of the speech and noise are estimated using a codebook-based approach, and the pitch parameters are estimated from the noisy signal using a harmonic model based approach \cite{nielsen2017fast}. 
We refer to this method in the rest of this paper as the Kalman-CB. This algorithm has been selected because of its good performance in noise reduction in terms of quality and intelligibility based on both objective and subjective measures.
The speech codebook was trained using 640 clean recordings selected from the MMPD data set (equally from both genders). 
To train the noise codebook, we used babble, restaurant, office and street noises to create four sub-codebooks. 
During the testing phase, all sub-codebooks, except the one corresponding to the target noise, were concatenated to form the final noise codebook.
The size of the speech and noise codebooks were set to 8 and 12, respectively.

The second supervised enhancement method is the DNN-based algorithm proposed in \cite{han2015learning}. 
This algorithm is the same as the one we used for dereverberation experiments, except it is trained using the noisy signals. 
This algorithm has been selected because, besides improvements in objective measures, it showed improvement in performance of automatic speech recognition in noisy environments.
To train the DNN, we used the same 640 clean recording that we used for training the speech codebook in the Kalman-CB algorithm.
The recordings were contaminated by three types of noise, namely babble, factory and F16 noises taken from NOISEX-92 database \cite{varga1993assessment} under different SNR conditions selected randomly from the continuous interval [0,10] dB. 

We used, as an unsupervised speech enhancement method, the algorithm proposed in \cite{erkelens2007minimum} which is based on the minimum mean-square error (MMSE) estimation of discrete Fourier transform (DFT) coefficients of speech while assuming a generalized gamma prior for the speech DFT coefficients. This method, denoted as MMSE-GGP, is a popular unsupervised algorithm which uses the MMSE-based tracker for noise power spectral density estimation.

Fig.~\ref{fig: noise_reduction_all} shows the PD detection performance in terms of~AUC for different noise types and SNR conditions. It can be observed from the figures that enhancing the degraded voice signals with the supervised methods in general improves the performance whereas the unsupervised method shows improvement only in the low SNR range and degrades the PD detection performance in higher SNR scenarios.
The low performance of the unsupervised algorithm can be due to the fact that noise statistics in this case is estimated using a method proposed in \cite{gerkmann2012unbiased} which has been designed for running speech rather than the sustained vowels. 
This observation is somewhat consistent with the statement in \cite{Vasquez}, which suggested that applying an unsupervised enhancement algorithm to the voiced segments results in a degradation in PD detection performance.


\input{figures/noise_reduction_plots.tex}

\subsection{Joint Noise Reduction and Dereverberation}
\label{ssec: jointnoise_red_dereverb}

In Sections \ref{ssec: dereverb} and \ref{ssec: noise_reduction}, we showed the impact of noise reduction and dereverberation when one of these degradations was present in the signal. 
However, in some cases, the recordings may be degraded simultaneously by reverberation and background noise. 
There have been methods proposed for joint noise reduction and dereverberation with access to multiple channels \cite{habets2008joint, kodrasi2016joint}. 
Since we have restricted ourselves to single channel enhancement methods, and motivated by the improvement in the PD detection performance as a result of using the DNN-SE algorithm for noise reduction and dereverberation, in this section, we investigate the effectiveness of this algorithm in performing joint noise reduction and dereverberation. 
In this case, the input to the DNN is the log-magnitude spectrum of the signal which is degraded by reverberation and background noise. For training the DNN model, the same 640 clean recordings that we used in the previous enhancement experiments were filtered with RIRs of different RT60s ranging from 400 ms to 1 s with 200 ms steps. 
Then, three types of noise, namely babble, factory and F16 noises (taken from NOISEX-92 database) were randomly added to the reverberant signals at different SNRs selected uniformly at random from the continuous interval [0,10]~dB.
Table \ref{tab:joint_enh_babble} summarizes the impact of joint noise reduction and dereverberation using the DNN-SE algorithm on the PD detection performance.
In this table, we have also included the cases of infinite SNR and zero RT60 to investigate the effect of the enhancement system when the clean recordings or the ones degraded by only noise or reverberation were processed by this algorithm. 
It can be observed for the case of babble noise that the DNN-SE improves the PD detection performance in most of the cases when reverberation and background noise coexist and in the cases where only noise is present. However, in the case of only reverberation, the DNN-SE shows improvement only in the cases where RT60 is 400 ms and above. It should be noted that the babble noise used for training and testing were taken from two different noise databases.
In the case of restaurant noise, improvement in PD detection performance is observed only in the low SNRs, namely -2 dB and -6 dB.
The results of the restaurant noise is interesting in a sense that it shows how the DNN-SE algorithm can generalize for a noise type not seen during the training phase. 

\input{tabels/joint_enhancement_babble.tex}

\section{Automatic Quality Control in Pathological Voice Recordings}
\label{sec: degradation_det}
We have shown in the previous section that, assuming the specific degradation is known, there exist algorithms to effectively transform a voice signal from a degraded condition into the acoustic condition in which models are trained. 
Choosing the appropriate enhancement algorithm, however, requires prior knowledge about the presence and type of degradation in a voice signal. In this section, we introduce two approaches to automatically control the quality of recordings. 
The first approach detects, at recording level, the presence and type of degradation which has influenced the majority of frames of the signal. The second approach, on the other hand, detects short-term degradations and protocol violations in a signal.

\subsection{Recording-Level Quality Control} \label{ssec:rec_level_deg_det}
The major limitation of the classification-based approaches for identifying the type of degradation in a voice signal \cite{Poorjam2017,poorjam2018parametric} is that they do not consider the fact that a recording can be subject to an infinite number of possible combinations of degradations in real scenarios. 
This causes some problems when a signal is contaminated by a new type of degradation for which the classifier has not been trained. 
Moreover, there is no control in class assignment for a high-quality outlier which do not comply with the context of the data set. 

To overcome these limitations, instead of using a multiclass classifier, we propose to use a set of parallel likelihood ratio detectors for the major types of degradations commonly encountered in remote voice analysis, each detecting a certain degradation type. 
This way, the likelihood ratio statistics of an observation given each of the models can be translated to the degree of contribution of each degradation to the degraded observation. Moreover, completely new degradation types and high-quality outliers can be detected if all models reject those observations according to a pre-defined threshold.

In this approach, the task of each detector is to determine whether a feature vector of the time frame $t$ of a voice signal, $\bm{x}_t$, was contaminated by the corresponding degradation, $H_0$, or not, $H_1$. 
The decision about the adherence of each frame of a given speech signal to the hypothesized degradation is then computed as:
\begin{equation}\label{eq:llk}
    \log p(\bm{x}_t|H_0) - \log p(\bm{x}_t|H_1) 
    \begin{cases}
    \geq\omega,& \text{accept}\ H_0\\
    <\omega,& \text{reject}\ H_0,
    \end{cases}
\end{equation}
where $\omega$ is a pre-defined threshold for detection, and $p(\bm{x}_t|H_0)$ and $p(\bm{x}_t|H_1)$ are respectively the likelihood of the hypotheses $H_0$ and $H_1$ given $\bm{x}_t$. 

To model the characteristics of each hypothesized degradation, we propose to fit a GMM of the likelihood function defined in (\ref{eq:gmm_lh_function}) to the frames of the recordings in the feature space. The motivation for using GMMs is that they are computationally efficient models that are capable of modeling sufficiently complex densities as a linear combination of simple Gaussians. Thus, the underlying acoustic classes of the signals might be modeled by individual Gaussian components. 
While the hypothesized degradation models can be well characterized by using training voice signals contaminated by the corresponding degradation, it is very challenging to model the alternative hypothesis as it should represent the entire space of all possible negative examples expected during recognition. 
To model the alternative hypothesis, instead of using individual degradation-specific alternative models, we train a single degradation-independent GMM using a large number of clean, degraded and outlier voice signals. 
Since this background model is used as an alternative hypothesis model for all hypothesized degradations, it is referred to as a universal background model~(UBM).  

When the UBM is trained, a set of degradation-dependent GMMs for modeling clean, noisy, reverberant and distorted recordings, $\mathcal{D}=\{\lambda_d\}_{d=1}^4$, are derived by adapting the parameters of the UBM through a \emph{maximum a posteriori} estimation and using the corresponding training data. 
Given the UBM, $\lambda_{\mathrm{ubm}}$, and the $d^{\mathrm{th}}$ trained degradation model, $\lambda_d$, and assuming that the feature vectors are independent, the log-likelihood ratio for a test observation, $\bm{X}_{\mathrm{ts}}=(\bm{x}_1,\ldots,\bm{x}_t,\ldots,\bm{x}_T)$, is calculated as:\\
\begin{equation}\label{eq:llk_dg}
    \sigma_d(\bm{X}_{\mathrm{ts}})=\frac{1}{T}\bigg(\sum_{t=1}^T\log p(\bm{x}_t|\lambda_{d})-\sum_{t=1}^T\log p(\bm{x}_t|\lambda_{\mathrm{ubm}})\bigg).
\end{equation}
The scaling factor in (\ref{eq:llk_dg}) is used to make the log-likelihood ratio independent of the signal duration and to compensate for the strong independence assumption for the feature vectors \cite{Reynolds2000}.
The decision for the test observation can be made by setting a threshold over the scores.

To parametrize the recordings, we propose to use mel-frequency cepstral coefficients (MFCCs) \cite{Deller2000}. Because it has been demonstrated in \cite{Poorjam2017,Poorjam2018} that degradation in speech signals predictably modifies the distribution of the MFCCs by changing the covariance of the features and shifting the mean to different regions in feature space, and the amount of change is related to the degradation level.

\subsubsection{Experimental Setup}\label{sssec:utt_deg_det_exp_setup}
For training the UBM, we randomly selected 8,000 recordings from the MMPD data set. To make the training data balanced over the subpopulations to avoid the model to be biased towards the dominant one, we randomly divided this subset into 5 equal partitions of 1,600 samples. The recordings of the first partition were randomly contaminated by six different types of noise namely babble, street, restaurant, office, white Gaussian and wind noises under different SNR conditions ranging from -10 dB to 20 dB in 2 dB steps. 
The recordings of the second partition were filtered by 46 real room impulse responses (RIRs) of the AIR database \cite{Jeub2009}, measured with mock-up phone in different realistic indoor environments, to produce reverberant data.
As an example of non-linearities in signals, the recordings of the third partition were processed randomly by either clipping, coding or clipping followed by coding.
The clipping level was set to 0.3, 0.5 and 0.7. We used 9.6 kbps and 16 kbps code-excited linear prediction (CELP) codecs \cite{Schroeder1985}. 
To consider the combination of degradations in signals, the recordings of the forth partition were randomly filtered by 46 different real RIRs and added to the noises typically present in indoor environments, namely babble, restaurant and office noise at 0 dB, 5 dB and 10 dB. The recordings of the last partition were used without any processing. The last subset also contains some outliers which do not contain relevant information for PD detection.

For adaptation of the degradation-dependent models, a subset of 800 good-quality recordings of PD patients and healthy speakers of both genders were equally selected from the MMPD data set. From this subset, 200 recordings were corrupted by babble, restaurant, street and office noises under different SNR conditions ranging from -5 dB to 10 dB in 5 dB steps. Another subset of 200 recordings were selected to be filtered by 16 real RIRs from AIR database. A subset of 200 recordings were also chosen to represent nonlinear distortions in signals by processing them in a same way the UBM data were distorted. The remaining 200 recordings were kept unchanged to represent the clean samples.

Using a Hamming window, recordings were segmented into frames of 30 ms with 10 ms overlap. For each frame of a signal, 12 MFCCs together with the log energy are calculated along with \emph{delta} and \emph{double-delta} coefficients. They are concatenated to form a 39-dimensional feature vector.

\subsubsection{Results}
To evaluate the proposed approach in identifying degradations in data not observed during the training phase, we used 10-fold cross validation with 10 iterations. 
For each experiment, we extended the test subset by adding 20 good-quality outlier recordings, including irrelevant sounds for PD detection randomly selected from the MMPD data set, to show whether the detectors could reject such outliers. 
Moreover, as an example of combination of degradations in speech signals, 20 good-quality recordings were selected from the MMPD data set, contaminated by noise and reveberation in a similar way we did for the UBM data, and appended them to the test subset to investigate whether both the noise and reverberation detectors could identify these recordings.

\begin{figure}[tb]
\centerline{\includegraphics[width=0.4\textwidth]{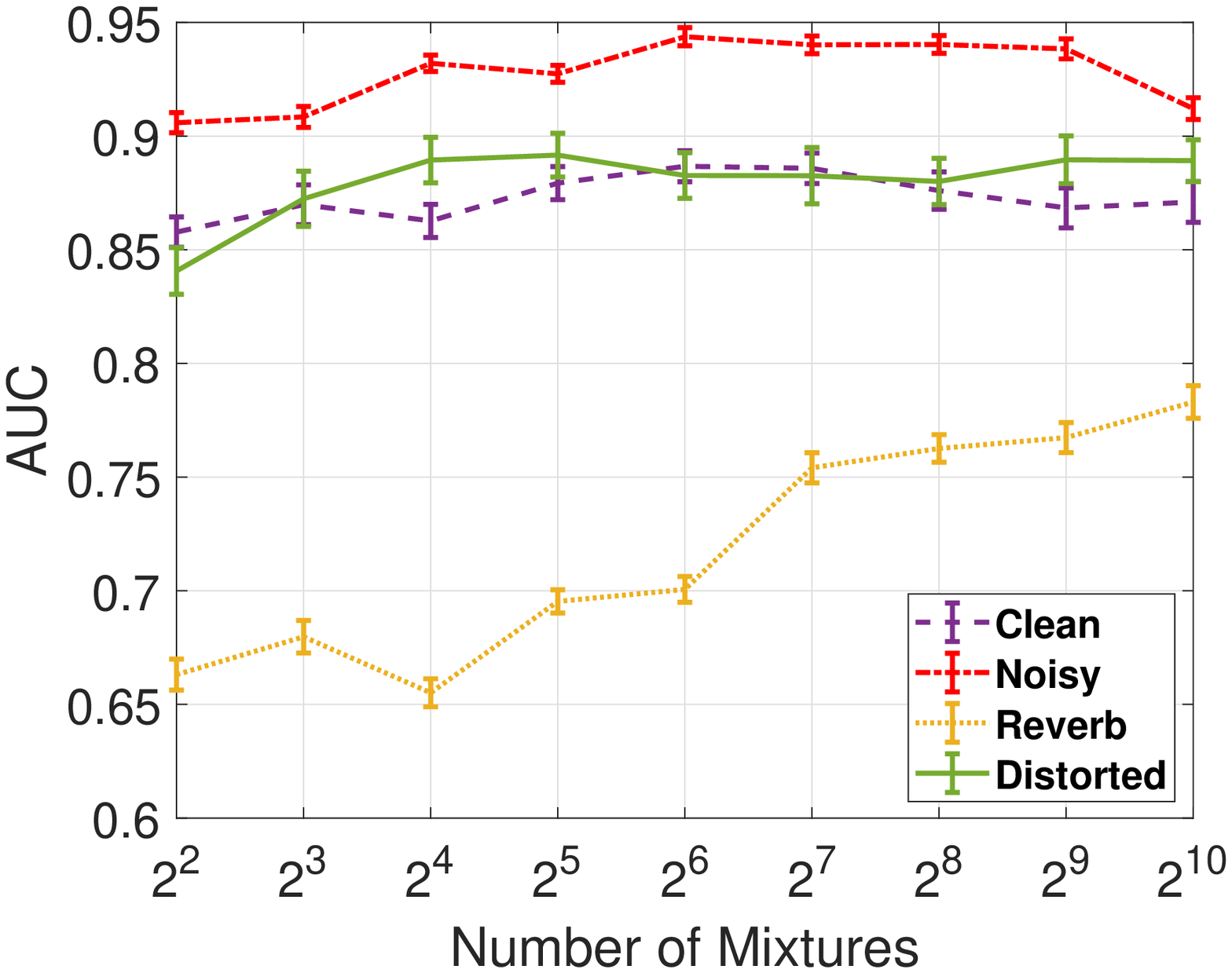}}
\vspace{-2mm}
\caption{The performance of the proposed recording-level degradation detection in terms of AUC, along with 95\% confidence intervals, as a function of number of mixture components.}
\label{fig:utt_lev_deg_det}
\end{figure}

Fig.~\ref{fig:utt_lev_deg_det} shows the performance of the detectors in terms of AUC, along with 95\% confidence intervals, as a function of the number of mixture components in GMMs. 
We can observe from the results that the degradations in voice signals are effectively identified when GMMs with 1024 mixtures are used. 
The lower performance for reverberation detection model is mainly due to misdetection of some of the recordings in which noise and reverberation coexist but the noise is more dominant than the reverberation.
This can also be explained by considering the analysis of vowels in the presence of different degradations \cite{Poorjam2017} which shows that MFCCs of the reverberant signals are, on average, positioned closer to the MFCCs of the clean signals, while noise and distortion (clipping) shift the MFCCs farther away from the position of clean MFCCs.

\subsection{Frame-Level Quality Control}
While many types of degradation, such as reverberation and nonlinear distortions, typically influence the entire recording, additive noise can have a short-term impact on a signal. 
Moreover, the test protocol can be violated for a short period of time in a remotely collected voice signal. 
In recording-level degradation detection, we assumed that the majority segments of a voice signal are influenced by some types of degradation. 
Likewise, if a voice sample is an outlier, the majority segments of the signal are assumed to contain irrelevant information for PD detection. Even though beneficial in providing a global information about the quality of a signal, it does not say whether a degraded or an outlier signal still contains useful segments to be considered for PD detection.
Identifying these segments facilitates making the most use of the available data.

In this paper, we consider additive noise as an example of a short-term degradation in a signal, and develop a framework which splits a voice signal into variable duration segments in an unsupervised manner by fitting an infinite hidden Markov model (iHMM) to the frames of the recordings in the MFCC domain. 
Then, the degraded segments and those that are associated with the protocol adherence or violation are identified by applying a multinomial naive Bayes classifier.

A HMM represents a probability distribution over sequences of observations $(\bm{x}_1,\ldots,\bm{x}_t,\ldots,\bm{x}_T)$ by invoking a Markov chain of hidden state variables $s_{1:T}=(s_1,\ldots,s_t,\ldots,s_T)$ where each $s_t$ is in one of the $K$ possible states \cite{Rabiner1989}. 
The likelihood of the observation $\bm{x}_t$ is modeled with a distribution of $K$ mixture components as:
\begin{equation}
    p(\bm{x}_t|s_{t-1}=i,\bm{\Theta})=\sum_{k=1}^K \pi_{i,k}p(\bm{x}_t|\bm{\theta}_k),
    \label{eq:hmm_llh}
\end{equation}
where $\bm{\Theta}=(\bm{\theta}_1,\ldots,\bm{\theta}_K)$ are the time-independent emission parameters, $\pi_{ij}=p(s_t=j|s_{t-1}=i)$, $(i,j=1,2,\ldots,K)$, is the transition matrix of $K\times K$.
We consider a HMM for clustering the frames of the signals in terms of different acoustic events. The prediction of the number of states required to cover all events such that we do not encounter unobserved events in the future is challenging. Moreover, it is reasonable to assume that as we observe more data, different types of protocol violations and acoustic events will appear and thus the inherent number of states will have to adapt accordingly.
Here, we propose to use an infinite HMM to relax the assumption of a fixed $K$ in (\ref{eq:hmm_llh}), which is defined as:
\begin{alignat}{4}
    &\bm{\beta} &&\sim &&\quad\mathrm{GEM}(\gamma)\nonumber\\
    &\bm{\pi}_k && \sim &&\quad\mathrm{DP}(\alpha,\bm{\beta}) &&\qquad (k=1,2,\ldots,\infty)\nonumber\\
    &\bm{\theta}_k &&\sim &&\quad H &&\qquad(k=1,2,\ldots,\infty)\nonumber \\
   &s_0 && = &&\quad 1\nonumber\\
   &s_t|s_{t-1} && \sim &&\quad \bm{\pi}_{s_{t-1}} &&\qquad(t=1,2,\ldots,T)\nonumber\\
    &\bm{x}_t|s_t && \sim &&\quad f(\bm{\theta}_{s_t})&&\qquad(t=1,2,\ldots,T).
\label{eq:ihmm}
\end{alignat} 
where $\bm{\pi}_k\sim\mathrm{DP}(\alpha,\bm{\beta})$ are drawn from a Dirichlet process (DP) with a local concentration parameter $\alpha>0$, $\bm{\beta}$ is the stick-breaking representation for DPs which is drawn from Griffiths-Engen-McCloskey (GEM) distribution with a global concentration parameter $\gamma>0$ \cite{Sethuraman1994}, each $\bm{\theta}_k$ is a sample drawn independently from the global base distribution over the component parameters of the HMM $H$, and $f$ is the observation model for each state.
The iHMM can possibly have countably infinite number of hidden states.
Using the direct assignment Gibbs sampler, which marginalizes out the infinitely many transition parameters, we infer the posterior over the sequence of hidden states  $\bm{\pi}$ and emission parameters $\bm{\Theta}$. In each iteration of the Gibbs sampling, we first re-sample the hidden states and then the base distribution parameters. For more details about the inference, we refer to \cite{Poorjam2019a}.

Considering an iHMM as a clustering algorithm, segments of the voice recordings with similar characteristics are clustered together under the same state indicator values. 
To identify the segments of the signal that are sufficiently reliable for detecting PD voice symptoms, those that need enhancement before being used for PD detection, and those which do not contain relevant information for PD detection, we propose to use the multinomial naive Bayes classifier to map the state indicators $s_{1:T}$ to the labels $y_{1:T}=(y_1,\ldots,y_t,\ldots,y_T)$, where $y_t=1$ if $\bm{x}_t$ adheres to the protocol, $y_t=2$ if it complies with the protocol but is degraded by additive noise, or $y_t=3$ if it violates the protocol.
In the multinomial naive Bayes, we assume that the samples in different classes have different multinomial distributions, and a feature vector for the $t^{\mathrm{th}}$ observation $\bm{\rho}_t=(\rho_{t,1},\ldots,\rho_{t,K})$ is a histogram, with $\rho_{t,k}$ being the number of times state $k$ is observed.
The likelihood of the histogram of a new observation $\tilde{\bm{\rho}}$ is defined as:
\begin{equation}
    P(\tilde{\bm{\rho}}|y_{1:T},\tilde{y},\bm{\rho}_{1:T})=\frac{(\sum_{k=1}^K \rho_{t,k})!}{\prod_{k=1}^K \rho_{t,k}!}\prod_{k=1}^K p_{k,\tilde{y}}^{\rho_{t,k}},
\end{equation}
where $p_{k,\tilde{y}}$ is the probability of the $k^{\mathrm{th}}$ attribute being in class $\tilde{y}\in\lbrace1,2,3\rbrace$, which is trained using the training data. 
Using the Bayes rule and the prior class probability $P(\tilde{y})$, the class label for a new test observation is predicted as:
\begin{equation}
    \hat{y}=\operatorname*{arg\,max}_{y\in\lbrace1,2,3\rbrace}\bigg(\log P(\tilde{y}=y)+\sum_{k=1}^K \tilde{\rho}_k \log(p_{k,y})\bigg).
\end{equation}

\subsubsection{Experimental Setup} \label{sec:experimental_setup}
To evaluate the performance of the proposed method, a subset of 100 good-quality recordings (50 PD patients and 50 healthy controls equally from both genders) has been selected from the MMPD data set. From this subset, 50 recordings were selected and 60\% of each signal were degraded by adding noise. We used babble, office, restaurant, street and wind noises, under different SNR conditions ranging from -5 dB to 10 dB in steps of 2.5 dB. 
In addition, 20 recordings from the MMPD data set containing several short- and long-term protocol violations were selected and added to the subset.

Using a Hamming window, recordings are segmented into frames of 30 ms with 10 ms overlap. For each frame of a signal, 12 MFCCs along with the log energy are calculated. The features of every five consecutive frames are averaged to smooth out the impact of articulation \cite{Poorjam2018}, and to prevent capturing very small changes in signal characteristics, which results in producing many uninterpretable states. 
Thus, each observation represents an averaged MFCCs of $\approx$100 ms of a signal.
For the iHMM, we use the conjugate normal-gamma prior over the Gaussian state parameters, set the hyper-parameters $\alpha$$=$$\gamma=10$, and run the inference for 150~iterations.

\subsubsection{Results}\label{sec:results}
The top plot in Fig.~\ref{fig:frm_lev_deg_det} shows a segment of 10 seconds duration selected from the data set. The segments of the signal which adhere to the test protocol and those that need enhancement are hand-labeled and shaded in green and pink, respectively.
Fitting the iHMM to the data, 49 different states were discovered in this particular subset. 
The middle plot in Fig.~\ref{fig:frm_lev_deg_det} illustrates the generated states in different colors.
To evaluate the performance of the proposed approach for data not observed during the training phase (i.e. out of sample), we used 10-fold CV and repeated the procedure 10 times. 
The results, presented in Table \ref{tab:confusion}, indicate that the proposed method can automatically identify short-term degradation and protocol violations in pathological voices with a 0.1 second resolution and high accuracy.
\input{tabels/confusion_matrix_frame_level.tex}

\input{figures/fig_frame_level_deg_detection.tex}

\subsection{Integrating Quality Control and Enhancement Algorithms}
\label{ssec: application}
The proposed quality control approaches can be integrated with the enhancement algorithms for cleaning-up the remotely collected signals before they are being processed by a PD detection system. In this section, we evaluate how this integration can lead to improvement in PD detection accuracy. 

The recording-level algorithm can be used in many different ways to provide information about the presence and type of degradation in a signal for an automatic clean-up process.
For example, one possible scenario could be to convert the parallel detectors to a multi-class classifier by calculating the maximum \emph{a posteriori} probability for a new observation. Then, the enhancement algorithm for which the observation has the highest degradation class probability will be applied.
Nevertheless, the advantage of the proposed method over the classification-based techniques is its capability to detect outlier recordings and those degraded by a new type of degradation. Thus, alternative approach could be to exploit the detectors to activate or bypass a set of enhancement blocks connected in series (e.g. noise reduction followed by dereverberation).
This scenario not only allows enhancement of a signal degraded by more than one degradation, but also prevents outliers to be processed by the PD detection system.
However, since there is no ground truth health status label for the outlier recordings, it is not possible to evaluate the performance of the PD detection system in the presence of outliers. 
For this reason, we considered a simple scenario in which the test subset only contains clean, noisy and reverberant recordings. 
Since there was no outlier in the test samples, the problem is simplified to a multi-class classification task.
For the experiment, we used the same 160 test recordings we used for the enhancement experiments. From this subset, 60 recordings were randomly selected and corrupted by restaurant, office and street noises under different SNR conditions ranging from -5 dB to 7 dB in 4 dB steps. Another 60 randomly chosen recordings were filtered by 16 real RIRs from AIR database. The enhancement algorithm used in this experiment is the DNN-SE. The model for noise reduction was trained using the noisy recordings and  the model used for dereverberation was trained using reverberant recordings. Table \ref{tab:utt_level_enhancement_results} shows the PD detection performance in terms of AUC for four different scenarios: (1) when no enhancement is applied to the recordings, (2) when the recordings, regardless of the presence and type of degradation, were processed randomly by either of the enhancement algorithms, (3) when recordings were enhanced by the enhancement model selected based on the estimated degradation labels, and (4) when the degraded recordings were enhanced based on the ground truth degradation labels. Comparing the results of the first and the second rows with those of the third and the forth rows suggests that applying appropriate enhancement algorithms to the degraded signals leads to an improvement in PD detection performance, and the level of improvement is related to the accuracy of the degradation detection system.
\input{tabels/recording_levl_enhancement.tex}

In the next experiment, we investigate how the proposed frame-level quality control method can improve the performance of PD detection.
To this aim, we randomly added babble, restaurant, office and street noises to all 160 test recordings at different SNRs ranging from -5 dB to 10 dB in 5 dB steps. However, for making a signal noisy, instead of adding a noise to the entire signal, we randomly corrupted 60\% frames of the signal. The enhancement algorithm used in this experiment is the Kalman-CB.
In Table \ref{tab:frame_level_enhancement_results}, we compare the PD detection performance for four different scenarios: (1) when no enhancement is applied to the recordings, (2) when the entire signals are enhanced, (3) when the signals are enhanced based on the predicted labels, and (4) when the signals are enhanced based on the ground truth labels.
\input{tabels/frame_levl_enhancement.tex}
For the last two scenarios, only the segments of the signals identified/labeled as degraded were enhanced. Moreover, we dropped the features of the frames identified as protocol violation.
Comparing the result of second scenario with the last two scenarios, we can observe the superiority of integrating the proposed frame-level quality control and the enhancement algorithm in dealing with short-term degradation and protocol violations in recordings.

\section{Conclusion}
\label{sec:conclusion} 
Additive noise, reverberation and nonlinear distortion are three types of degradation typically encountered during remote voice analysis which cause an acoustic mismatch between training and operation conditions. In this paper, we investigated the impact of these degradations on the performance of a PD detection system.
Then, given that the specific degradation is known, we explored the effectiveness of a variety of the state-of-the-art enhancement algorithms in reducing this mismatch and, consequently, in improving the PD detection performance.
We showed how applying appropriate enhancement algorithms can effectively improve the PD detection accuracy.
To inform the choice of enhancement method, we proposed two quality control techniques operating at recording- and frame-level. The recording-level approach provides information about the presence and type of degradation in voice signals. The frame-level algorithm, on the other hand, identifies the short-term degradations and protocol violations in voice recordings.
Experimental results showed the effectiveness of the quality control approaches in choosing appropriate signal enhancement algorithms which resulted in improvement in the PD detection accuracy.



This study has important implications that extend well beyond the PD detection system. It can be considered as a step towards the design of robust speech-based applications capable of operating in a variety of acoustic environments. 
For example, since the proposed quality control approaches are not limited to specific speech types, they can be used as a pre-processing step for many end-to-end speech-based systems, such as automatic speech recognition, to make them more robust against different acoustic conditions. They might also be utilized to automatically control the quality of recordings in large-scale speech data sets.
Moreover, these approaches have the potential to be used for other sensor modalities to identify short- and long-term degradations and abnormalities which can help to choose an adequate action.


\bibliographystyle{./package_style/IEEE_tran} 
\bibliography{./ref/library.bib}

\end{document}

%% file: figures/degradations.tex
\begin{figure*}
    \centering
    \begin{subfigure}[b]{0.32\textwidth}
        \includegraphics[width=\textwidth]{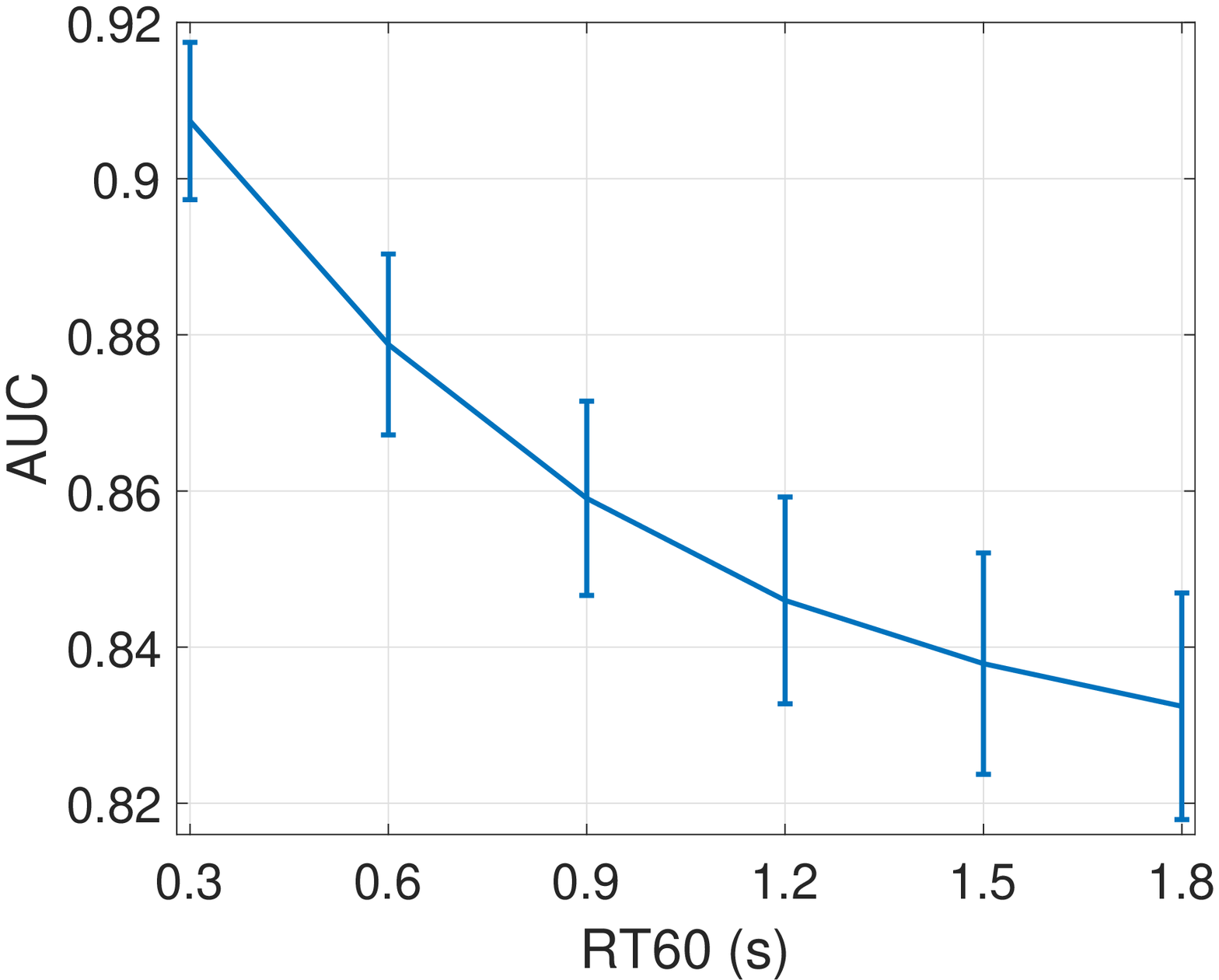}
        \caption{Reverberation}
        \label{fig: auc_exp1_RT60}
    \end{subfigure}
    \begin{subfigure}[b]{0.32\textwidth}
        \includegraphics[width=\textwidth]{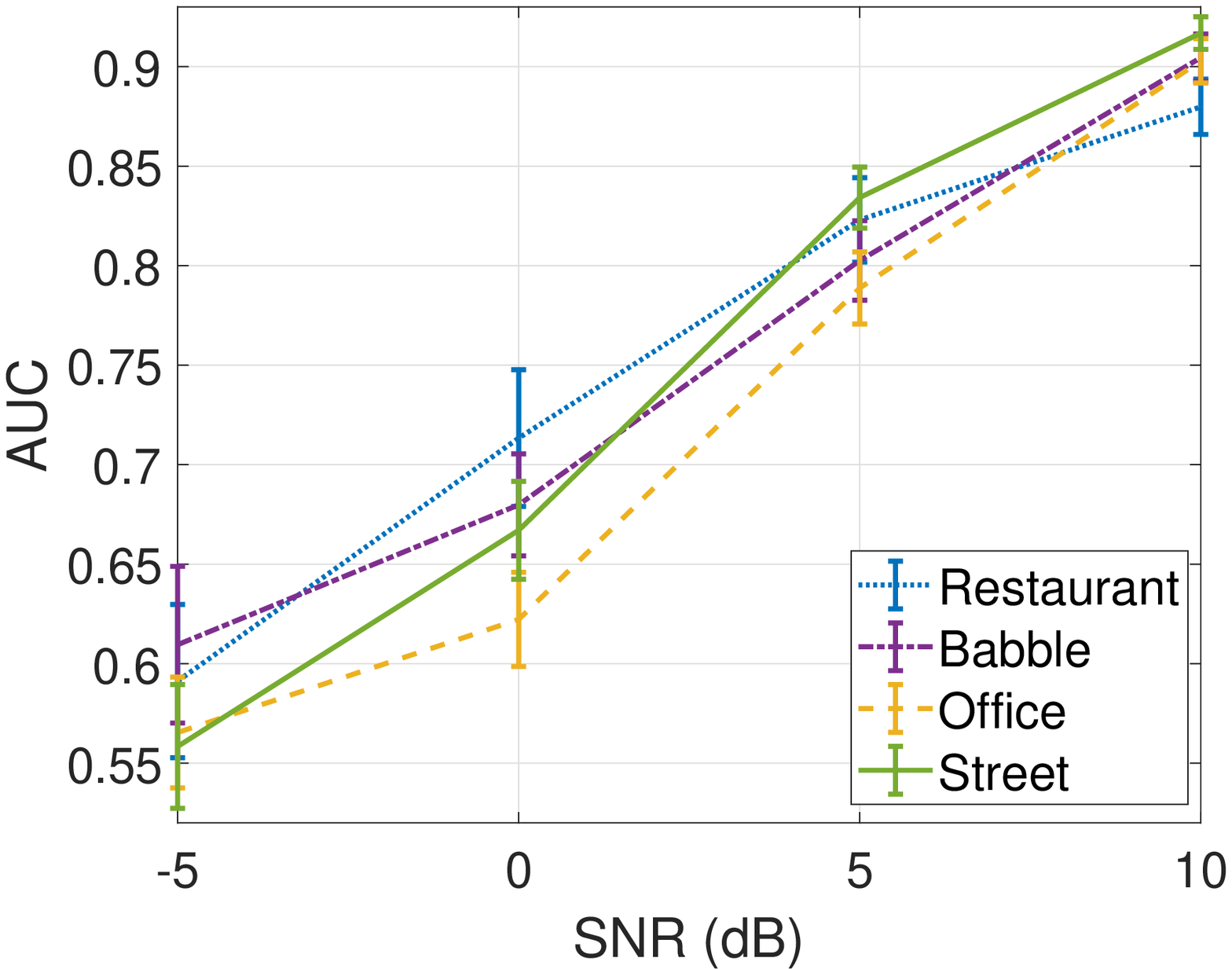}
        \caption{Additive Noise}
        \label{fig: auc_exp1_SNR}
    \end{subfigure}
    \begin{subfigure}[b]{0.32\textwidth}
        \includegraphics[width=\textwidth]{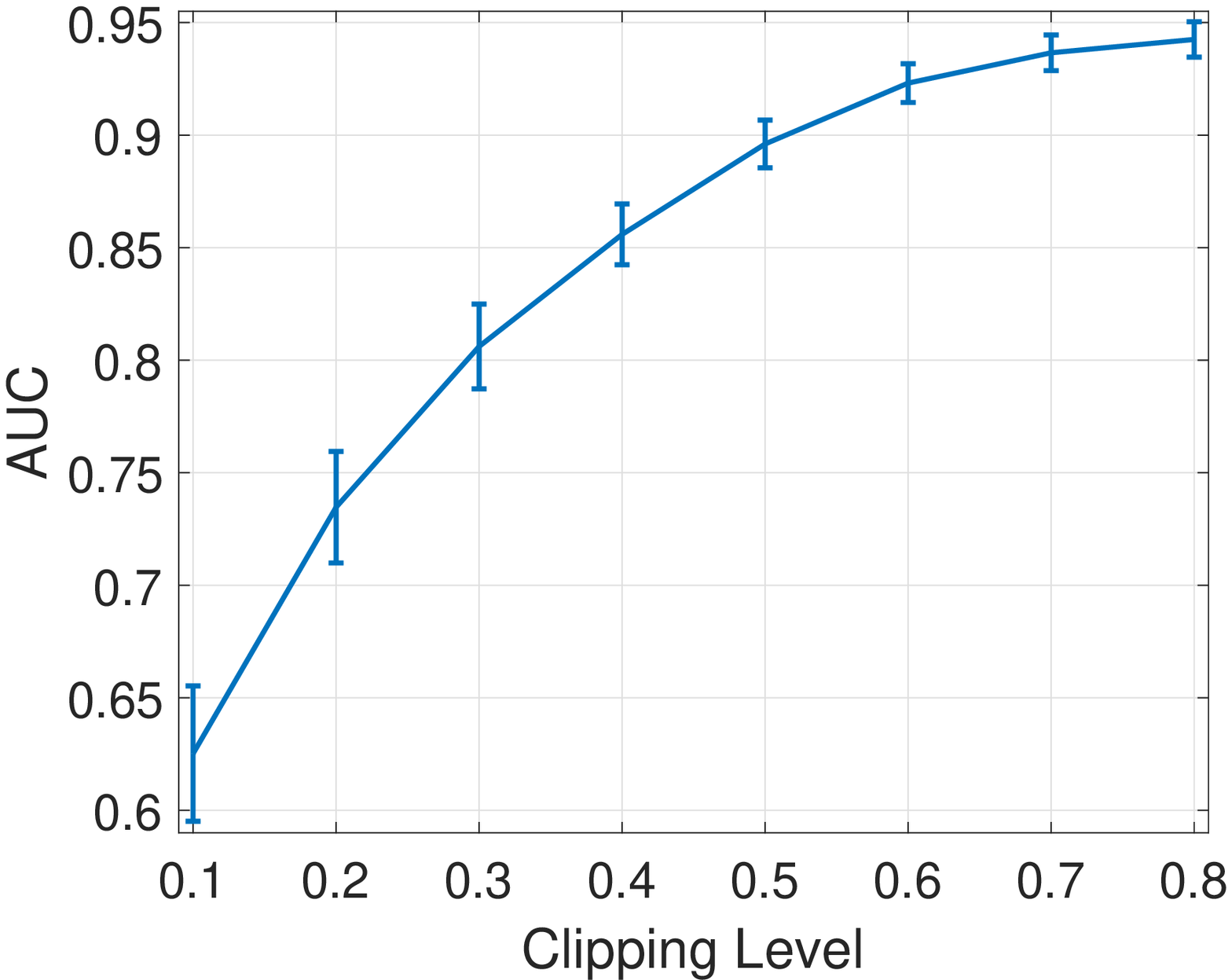}
        \caption{Nonlinear Distortion}
        \label{fig: auc_exp1_CLIP}
    \end{subfigure}
    \vspace{-2mm}
    \caption{Performance of the PD detection system in acoustic mismatch conditions due to different degradations in test signals in terms of AUC, along with 95\% confidence intervals.}\label{fig: degradation_all}
\end{figure*}

%% file: figures/noise_reduction_plots.tex
\begin{figure*}
    \centering
    \begin{subfigure}[b]{0.246\textwidth}
        \includegraphics[width=\textwidth]{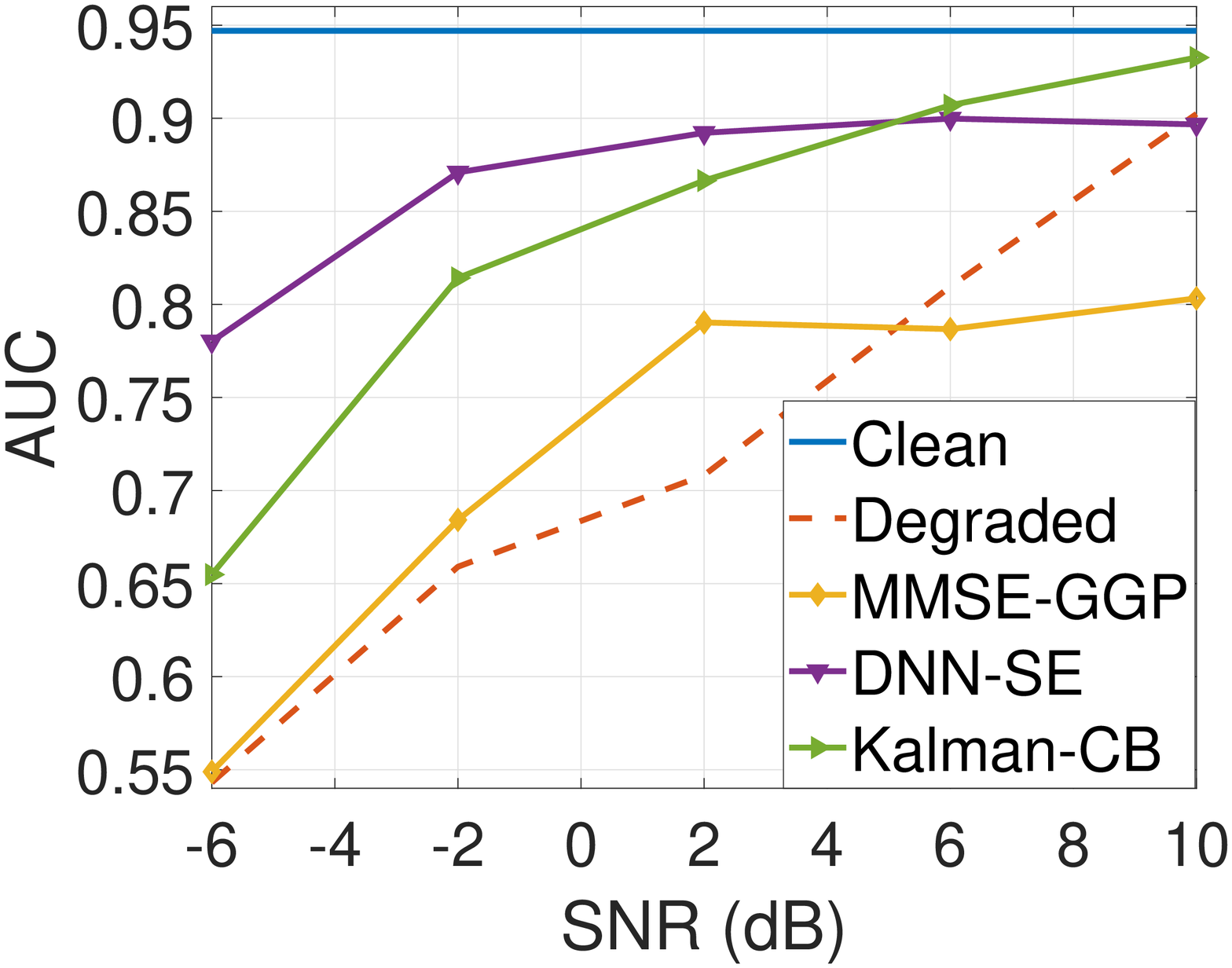}
        \caption{Babble Noise}
        \label{fig: auc_BB}
    \end{subfigure}
    \begin{subfigure}[b]{0.246\textwidth}
        \includegraphics[width=\textwidth]{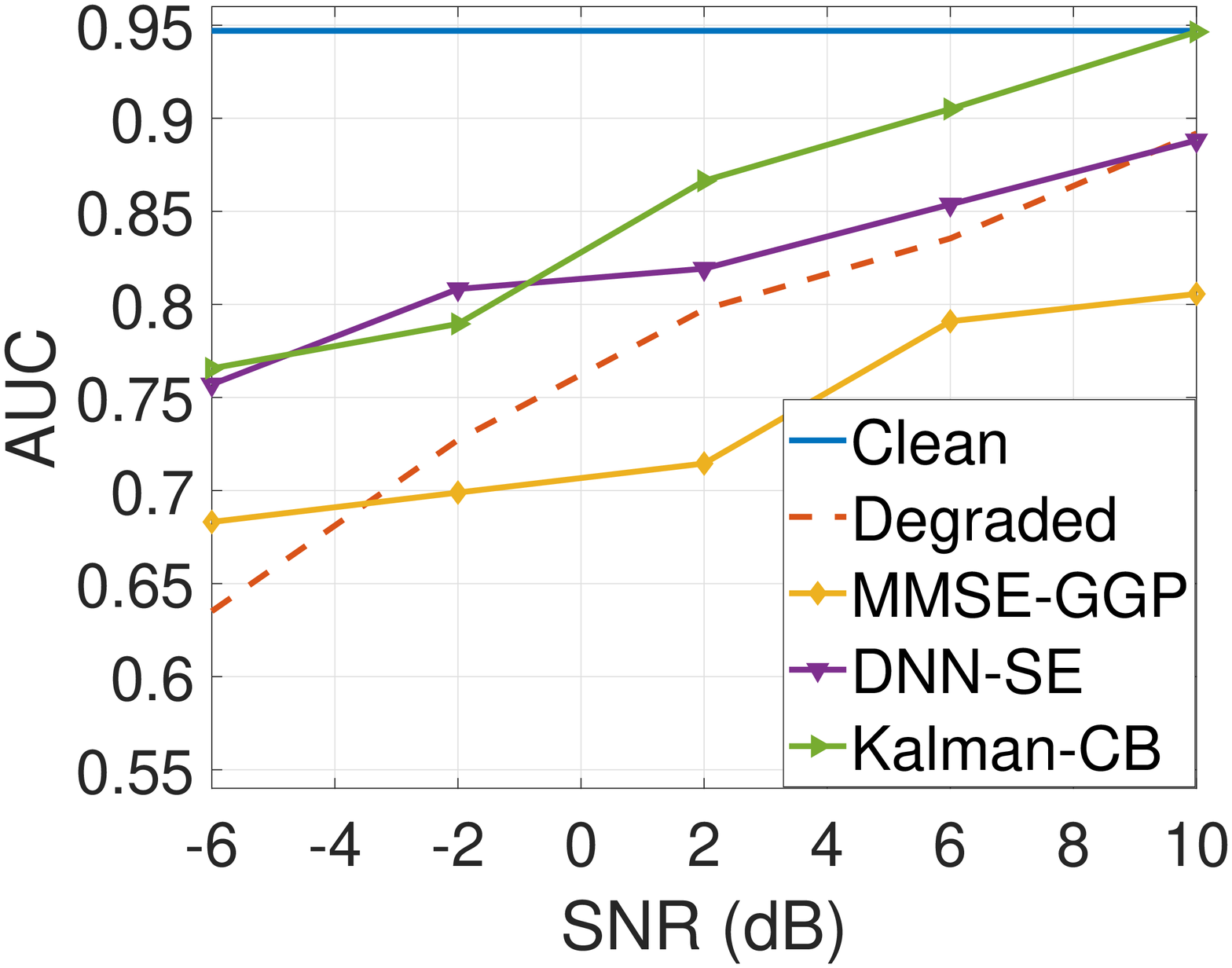}
        \caption{Restaurant Noise}
        \label{fig: auc_RS}
    \end{subfigure}
    \begin{subfigure}[b]{0.245\textwidth}
        \includegraphics[width=\textwidth]{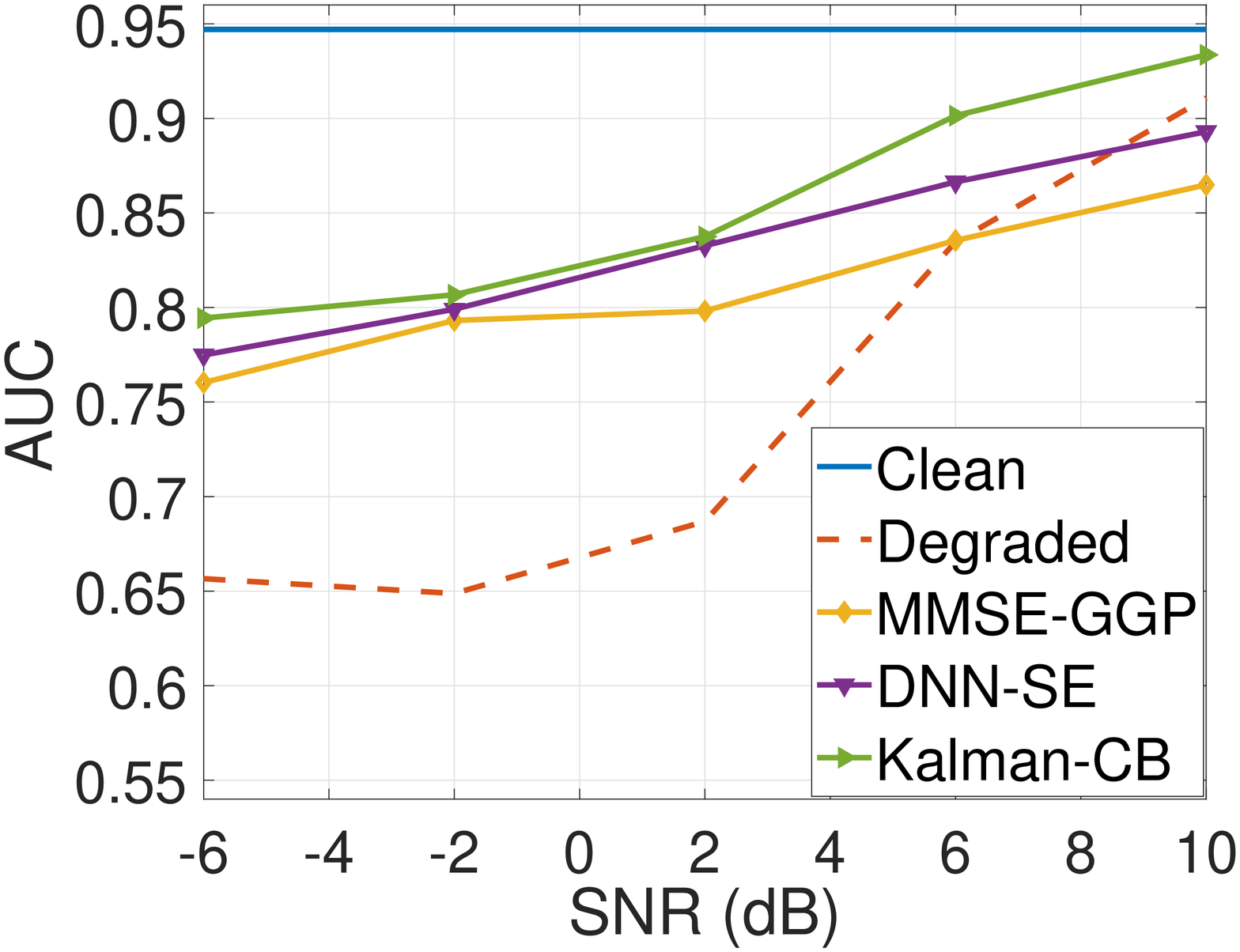}
        \caption{Office Noise}
        \label{fig: auc_OF}
    \end{subfigure}
    \begin{subfigure}[b]{0.245\textwidth}
        \includegraphics[width=\textwidth]{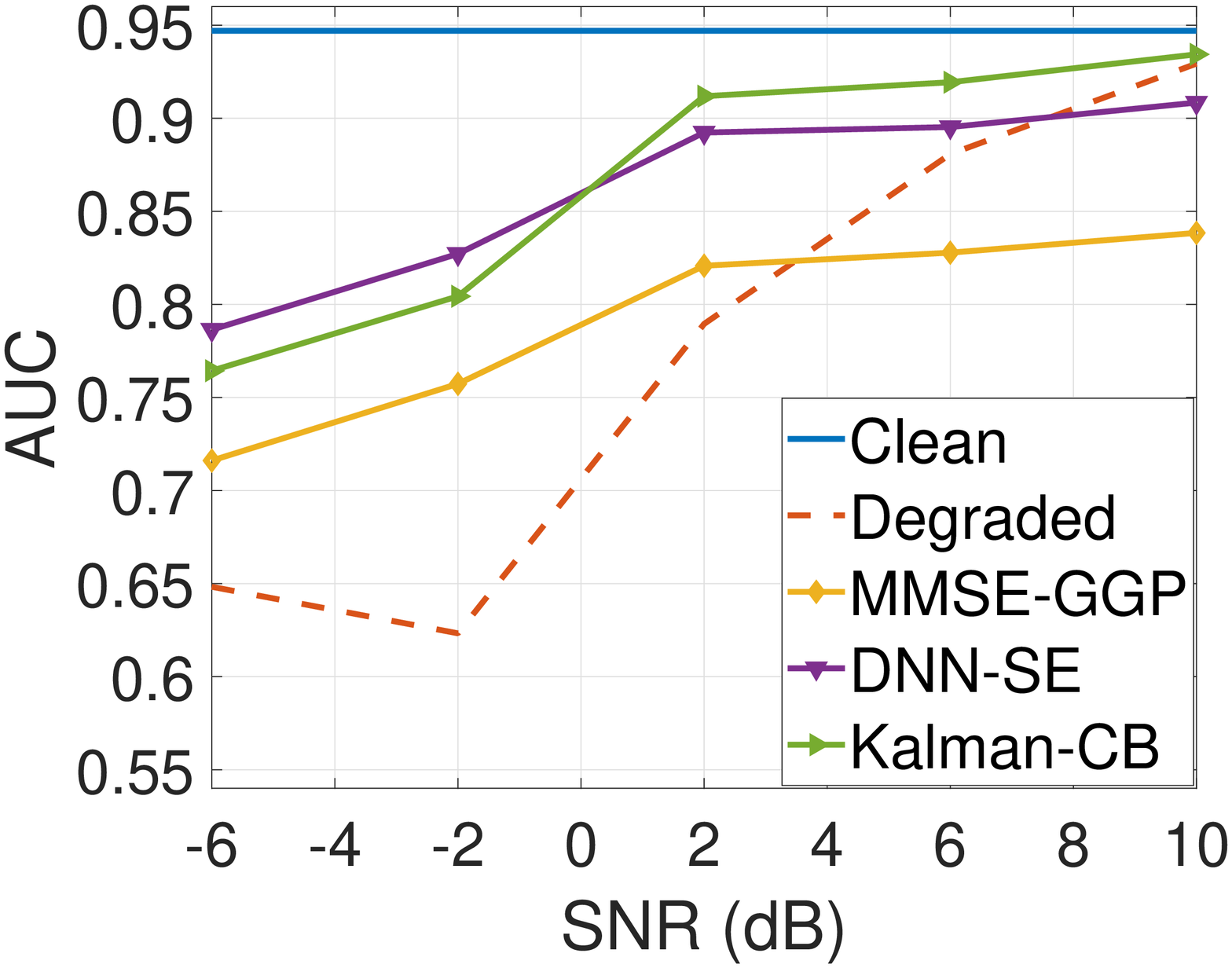}
        \caption{Street Noise}
        \label{fig: auc_ST}
    \end{subfigure}
    \caption{Impact of different noise reduction algorithms on the PD detection performance, in terms of AUC, under different noise types and SNR conditions.}
\label{fig: noise_reduction_all}
\end{figure*}

%% file: tabels/joint_enhancement_babble.tex
\begin{table*}[tb]
\caption{\label{tab:joint_enh_babble} {Impact of joint noise reduction and dereverberation using the DNN-SE algorithm on the PD detection performance. Bold numbers indicate the improvement in performance.}}
\renewcommand{\arraystretch}{1.2}
\centerline{
\begin{tabular}{|l|c|c|c|c|c|c|c|c||c|c|c|c|c|c|}
\cline{4-15}
\multicolumn{1}{c}{}&\multicolumn{1}{c}{} &\multicolumn{1}{c|}{} & \multicolumn{6}{c||}{\textbf{Babble Noise: SNR (dB) }} &  \multicolumn{6}{c|}{\textbf{Restaurant Noise: SNR (dB) }}   \\
\cline{4-15}
\multicolumn{1}{c}{} & \multicolumn{1}{c}{} &\multicolumn{1}{c|}{} & -6  & -2  & 2    & 6    & 10   & inf & -6  & -2  & 2    & 6    & 10   & inf  \\ 
\cline{1-15} \\ [-1.2em]  \cline{2-15}
\multirow{12}{*}{\begin{turn}{90}\textbf{RT60 (s)}\end{turn}} & \multirow{2}{*}{0} & Degraded  & 0.67 & 0.59 & 0.69 & 0.80 & 0.90 & 0.95 & 0.71 & 0.81 & 0.82 & 0.82 & 0.90 &0.95\\ \cline{3-15} 
                           & & DNN-SE & \textbf{0.80} & \textbf{0.89} & \textbf{0.89} & \textbf{0.89} & 0.89 & 0.91 & \textbf{0.77} & 0.81 & 0.82 & \textbf{0.83} & 0.87 & 0.91  \\ \cline{2-15} \\[-1.2em] \cline{2-15} 
                           & \multirow{2}{*}{0.2}                      & Degraded  & 0.56 & 0.64 & 0.72 & 0.81 & 0.89 & 0.95 & 0.67 & 0.75 & 0.76 & 0.85 & 0.89 & 0.95 \\ \cline{3-15} 
                           & & DNN-SE & \textbf{0.82} & \textbf{0.89} & \textbf{0.87} & \textbf{0.89} & 0.89 & 0.91 & \textbf{0.74} & \textbf{0.79} & \textbf{0.79} & 0.84 & 0.87 & 0.91 \\ \cline{2-15} \\[-1.2em] \cline{2-15} 
                           & \multirow{2}{*}{0.4} & Degraded  & 0.54 & 0.66 & 0.70 & 0.80 & 0.84 & 0.90& 0.62 & 0.73  & 0.83 &  0.83 & 0.83 & 0.92  \\ \cline{3-15} 
                           & & DNN-SE & \textbf{0.78} & \textbf{0.84} & \textbf{0.85} & \textbf{0.89} & \textbf{0.86} & \textbf{0.91} & \textbf{0.73} & \textbf{0.77} & 0.79 & 0.83 & 0.82 & 0.91\\ \cline{2-15} \\[-1.2em] \cline{2-15} 
                           & \multirow{2}{*}{0.6} & Degraded  & 0.64 & 0.70 & 0.71 & 0.78 & 0.81 & 0.88 &0.59 & 0.79 & 0.81 & 0.80 & 0.86 & 0.89\\ \cline{3-15} 
                           & & DNN-SE & \textbf{0.75} & \textbf{0.83} & \textbf{0.85} & \textbf{0.85} & \textbf{0.88} & \textbf{0.89} & \textbf{0.69} & \textbf{0.81} & 0.79 & \textbf{0.81} & 0.84 & \textbf{0.91}\\ \cline{2-15} \\[-1.2em] \cline{2-15} 
                           & \multicolumn{1}{l|}{\multirow{2}{*}{0.8}} & Degraded  & 0.67 & 0.70 & 0.73 & 0.79 & 0.83 & 0.89 & 0.58 & 0.76 & 0.82 & 0.81 & 0.86 & 0.87\\ \cline{3-15} & \multicolumn{1}{l|}{} & DNN-SE & \textbf{0.81} & \textbf{0.83} & \textbf{0.86} & \textbf{0.87} & \textbf{0.88} & \textbf{0.91} & \textbf{0.75} & 0.76 & 0.80 & \textbf{0.84} & \textbf{0.87} & \textbf{0.90} \\ \cline{2-15} \\[-1.2em] \cline{2-15} 
                           & \multirow{2}{*}{1} & Degraded  & 0.54 & 0.68 & 0.74 & 0.81 & 0.84 & 0.88 &  0.65 & 0.75 & 0.76 & 0.82 & 0.83 & 0.85 \\ \cline{3-15} 
                           & & DNN-SE & \textbf{0.80} & \textbf{0.81} & \textbf{0.86} & \textbf{0.86} & \textbf{0.88} & \textbf{0.90} & \textbf{0.76} & 0.75 & \textbf{0.78} & 0.81 & 0.82 & \textbf{0.90} \\ \hline
\end{tabular}}
\renewcommand{\arraystretch}{1}
\end{table*}

%% file: tabels/confusion_matrix_frame_level.tex
\begin{table}[tb]
\caption{\label{tab:confusion} {The confusion matrix of the proposed frame-level quality control method. Results are in the form of mean$\pm$STD.}}
\centerline{
\renewcommand{\arraystretch}{1.3}
\begin{tabular}{lc|c|c|c|}
 \cline{3-5} 
 & &\multicolumn{3}{c|}{\textbf{Predicted}}\\ 
\cline{3-5}
&&Adherence&Degraded&Violation\\ 
& & & &\\[-1.5em]
\hline 
\multicolumn{1}{|c|}{\multirow{3}{*}{\begin{turn}{90}\textbf{Actual \ }\end{turn}}} & \multicolumn{1}{c|}{Adherence} &$95\%\pm 1\%$& $3\%\pm 1\%$& $2\%\pm 1\%$\\
\cline{2-5}
\multicolumn{1}{|c|}{} & \multicolumn{1}{c|}{Degraded} & $10\%\pm 2\%$ & $89\%\pm 2\%$& $0\%\pm 0\%$ \\
\cline{2-5}
\multicolumn{1}{|c|}{} & \multicolumn{1}{c|}{Violation} & $5\%\pm 2\%$ & $2\%\pm 1\%$& $93\%\pm 2\%$ \\\hline 
\end{tabular}}
\renewcommand{\arraystretch}{1}
\vspace{-1mm}
\end{table}


%% file: figures/fig_frame_level_deg_detection.tex
\begin{figure}[tb]
\centerline{
\begin{overpic}[width=0.48\textwidth,height=0.22\textheight, tics=2,trim = 33mm 87mm 32mm 3mm, clip]{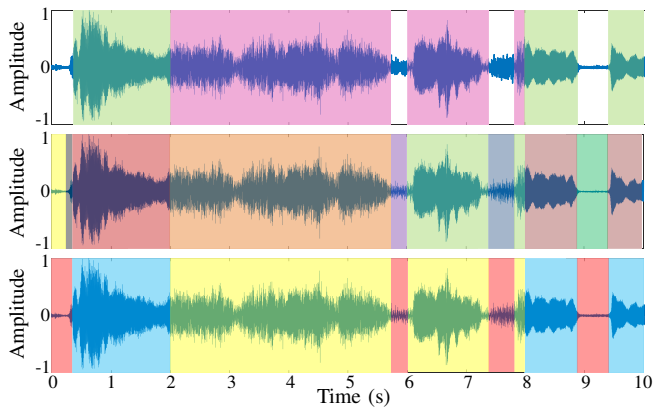}
\put (0,44) {\begin{turn}{90}\footnotesize{Amplitude}\end{turn}}
\put (0,24.7) {\begin{turn}{90}\footnotesize{Amplitude}\end{turn}}
\put (0,6.2) {\begin{turn}{90}\footnotesize{Amplitude}\end{turn}}
\put (4,4) {\scriptsize{-1}}
\put (4,23) {\scriptsize{-1}}
\put (4,42) {\scriptsize{-1}}
\put (4.8,12) {\scriptsize{0}}
\put (4.8,30.9) {\scriptsize{0}}
\put (4.8,49.9) {\scriptsize{0}}
\put (4.8,20.3) {\scriptsize{1}}
\put (4.8,39) {\scriptsize{1}}
\put (4.8,58) {\scriptsize{1}}
\put (47,-0.6) {\footnotesize{Time (s)}}
\put (6,1.8) {\scriptsize{0}}
\put (14.9,1.8) {\scriptsize{1}}
\put (24,1.8) {\scriptsize{2}}
\put (33.3,1.8) {\scriptsize{3}}
\put (42.3,1.8) {\scriptsize{4}}
\put (51.4,1.8) {\scriptsize{5}}
\put (60.4,1.8) {\scriptsize{6}}
\put (69.4,1.8) {\scriptsize{7}}
\put (78.4,1.8) {\scriptsize{8}}
\put (87.4,1.8) {\scriptsize{9}}
\put (95.4,1.8) {\scriptsize{10}}
\end{overpic}}
\vspace{-1mm}
\caption{Illustrative results of applying the proposed frame-level degradation detection method to a 10-second segment of the voice recordings selected from the data set. In the top plot, the green shaded and pink shaded areas represent the segments of the signal which are hand-labeled as adhering to the protocol and those need enhancement, respectively. 
The middle plot shows the states, generated by the iHMM, in different colors. 
The bottom plot illustrates the result of applying a trained classifier to the state indicators to predict which segments adhere to (shaded in blue), which ones violate the protocol (shaded in red), and which ones need enhancement (shaded in yellow).}
\vspace{-1mm}
\label{fig:frm_lev_deg_det}
\end{figure}

%% file: tabels/recording_levl_enhancement.tex
\begin{table}[tb]
\caption{\label{tab:utt_level_enhancement_results} {Evaluation of the impact of applying the proposed Recording-level quality control in combination with DNN-SE on the PD detection performance.}}
\vspace{-1.5mm}
\centerline{ %
\begin{tabular}{|l|c|}
\hline 
{\footnotesize\textbf{Scenarios}} & {\footnotesize\textbf{AUC}} 
\tabularnewline
\hline 
\hline 
{\footnotesize\textbf{No Enhancement}} & {\footnotesize {0.84}} \tabularnewline
\hline 
{\footnotesize\textbf{Enhancement based on Randomly Chosen Algorithm}} & {\footnotesize {0.86}} \tabularnewline
\hline 
{\footnotesize\textbf{Enhancement based on Predicted Labels}} & {\footnotesize {0.89}} \tabularnewline
\hline 
{\footnotesize\textbf{Enhancement based on Ground Truth Labels}} & {\footnotesize {0.90}} \tabularnewline
\hline 
\end{tabular}}
\end{table}

%% file: tabels/frame_levl_enhancement.tex
\begin{table}[tb]
\caption{\label{tab:frame_level_enhancement_results} {Evaluation of the impact of applying the proposed frame-level quality control on the PD detection performance.}}
\vspace{-1.5mm}
\centerline{ %
\begin{tabular}{|l|c|}
\hline 
{\footnotesize\textbf{Scenarios}} & {\footnotesize\textbf{AUC}} 
\tabularnewline
\hline 
\hline 
{\footnotesize\textbf{No Enhancement}} & {\footnotesize {0.86}} \tabularnewline
\hline 
{\footnotesize\textbf{Enhancement of Entire Recording}} & {\footnotesize {0.89}} \tabularnewline
\hline 
{\footnotesize\textbf{Enhancement based on Predicted Labels}} & {\footnotesize {0.92}} \tabularnewline
\hline 
{\footnotesize\textbf{Enhancement based on Ground Truth Labels}} & {\footnotesize {0.93}} \tabularnewline
\hline 
\end{tabular}}
\end{table}

%% file: QC_SE_PD.bbl
\begin{thebibliography}{10}
\providecommand{\url}[1]{#1}
\csname url@samestyle\endcsname
\providecommand{\newblock}{\relax}
\providecommand{\bibinfo}[2]{#2}
\providecommand{\BIBentrySTDinterwordspacing}{\spaceskip=0pt\relax}
\providecommand{\BIBentryALTinterwordstretchfactor}{4}
\providecommand{\BIBentryALTinterwordspacing}{\spaceskip=\fontdimen2\font plus
\BIBentryALTinterwordstretchfactor\fontdimen3\font minus
  \fontdimen4\font\relax}
\providecommand{\BIBforeignlanguage}[2]{{%
\expandafter\ifx\csname l@#1\endcsname\relax
\typeout{** WARNING: IEEEtran.bst: No hyphenation pattern has been}%
\typeout{** loaded for the language `#1'. Using the pattern for}%
\typeout{** the default language instead.}%
\else
\language=\csname l@#1\endcsname
\fi
#2}}
\providecommand{\BIBdecl}{\relax}
\BIBdecl

\bibitem{Ishihara2007}
L.~S. Ishihara, A.~Cheesbrough, C.~Brayne, and A.~Schrag, ``{Estimated life
  expectancy of Parkinson's patients compared with the UK population},''
  \emph{Journal of Neurol Neurosurg Psychiatry}, vol.~78, pp. 1304--1309, 2007.

\bibitem{Ho1998}
A.~K. Ho, R.~Iansek, C.~Marigliani, J.~L. Bradshaw, and S.~Gates, ``{Speech
  impairment in a large sample of patients with Parkinson's disease.}''
  \emph{Behavioural Neurology}, vol.~11, no.~3, pp. 131--137, 1998.

\bibitem{Eliasova}
I.~Eliasova, J.~Mekyska, M.~Kostalova, R.~Marecek, Z.~Smekal, and I.~Rektorova,
  ``{Acoustic evaluation of short-term effects of repetitive transcranial
  magnetic stimulation on motor aspects of speech in Parkinson's disease},''
  \emph{Journal of Neural Transmission}, vol. 120, no.~4, pp. 597--605, 2013.

\bibitem{Tsanas2012}
A.~Tsanas, M.~A. Little, P.~E. McSharry, J.~Spielman, and L.~O. Ramig, ``{Novel
  speech signal processing algorithms for high-accuracy classification of
  Parkinson's disease},'' \emph{IEEE Transactions on Biomedical Engineering},
  vol.~59, pp. 1264--1271, 2012.

\bibitem{Zhan2016}
A.~Zhan, M.~A. Little, D.~A. Harris, S.~O. Abiola, E.~R. Dorsey, S.~Saria, and
  A.~Terzis, ``{High frequency remote monitoring of Parkinson's disease via
  smartphone: platform overview and medication response detection},''
  \emph{arXiv preprint arXiv:1601.00960}, pp. 1--12, 2016.

\bibitem{Gil2006}
D.~Gil and M.~Johnson, ``{Diagnosing Parkinson by using artificial neural
  networks and support vector machines},'' \emph{Global Journal of Computer
  Science and Technology}, pp. 63--71, 2009.

\bibitem{Rusz2018}
J.~Rusz, J.~Hlavni{\v{c}}ka, T.~Tykalov{\'a}, M.~Novotn{\'y}, P.~Du{\v{s}}ek,
  K.~{\v{S}}onka, and E.~Ru{\v{z}}i{\v{c}}ka, ``{Smartphone allows capture of
  speech abnormalities associated with high risk of developing Parkinson's
  disease},'' \emph{IEEE Transactions on Neural Systems and Rehabilitation
  Engineering}, vol.~26, no.~8, pp. 1495--1507, 2018.

\bibitem{Fan2014}
J.~Fan, F.~Han, and H.~Liu, ``{Challenges of big data analysis},''
  \emph{National Science Review}, vol.~1, no.~2, pp. 293--314, 2014.

\bibitem{Gong1995}
Y.~Gong, ``{Speech recognition in noisy environments: A survey},'' \emph{Speech
  Communication}, vol.~16, no.~3, pp. 261--291, 1995.

\bibitem{Fakhry2018}
M.~Fakhry, A.~H. Poorjam, and M.~G. Christensen, ``{Speech enhancement by
  classification of noisy signals decomposed using NMF and Wiener filtering},''
  in \emph{26th European Signal Processing Conference}, 2018.

\bibitem{Hansen2014}
J.~H.~L. Hansen, A.~Kumar, and P.~Angkititrakul, ``{Environment mismatch
  compensation using average eigenspace-based methods for robust speech
  recognition},'' \emph{International Journal of Speech Technology}, vol.~17,
  no.~4, pp. 353--364, 2014.

\bibitem{Alam2017}
J.~Alam, P.~Kenny, G.~Bhattacharya, and M.~Kockmann, ``{Speaker verification
  under adverse conditions using i-vector adaptation and neural networks},'' in
  \emph{Interspeech}, 2017, pp. 3732--3736.

\bibitem{Mammone1996a}
R.~J. Mammone, {Xiaoyu Zhang}, and R.~P. Ramachandran, ``{Robust speaker
  recognition: A feature-based approach},'' \emph{IEEE Signal Processing
  Magazine}, vol.~13, no.~5, p.~58, 1996.

\bibitem{Nercessian2016}
S.~Nercessian, P.~Torres-Carrasquillo, and G.~Martinez-Montes, ``{Approaches
  for language identification in mismatched environments},'' in \emph{IEEE
  Spoken Language Technology Workshop}, 2016, pp. 335--340.

\bibitem{Poorjam2016}
A.~H. Poorjam, R.~Saeidi, T.~Kinnunen, and V.~Hautam{\"{a}}ki, ``{Incorporating
  uncertainty as a quality measure in i-vector based language recognition},''
  in \emph{Speaker and Language Recognition Workshop}, Bilbao, Spain, 2016, pp.
  74--80.

\bibitem{Vasquez}
J.~Vasquez-Correa, T.~Arias-Vergara, J.~R. Orozco-Arroyave, J.~F.
  Vargas-Bonilla, J.~D. Arias-Londono, and E.~N{\"{o}}th, ``Automatic detection
  of {Parkinson's} disease from continuous speech recorded in real-world
  conditions,'' in \emph{Interspeech}, 2015, pp. 3--7.

\bibitem{Poorjam2019}
A.~H. Poorjam, M.~A. Little, J.~R. Jensen, and M.~G. Christensen, ``{Quality
  control in remote speech data collection},'' \emph{IEEE Journal of Selected
  Topics in Signal Processing}, vol.~13, no.~2, 2019.

\bibitem{Poorjam2017}
A.~H. Poorjam, J.~R. Jensen, M.~A. Little, and M.~G. Christensen, ``{Dominant
  distortion classification for pre-processing of vowels in remote biomedical
  voice analysis},'' in \emph{Interspeech}, 2017, pp. 289--293.

\bibitem{poorjam2018parametric}
A.~H. Poorjam, M.~A. Little, J.~R. Jensen, and M.~G. Christensen, ``A
  parametric approach for classification of distortions in pathological
  voices,'' in \emph{IEEE International Conference on Acoustics, Speech and
  Signal Processing}, 2018, pp. 286--290.

\bibitem{Badawy2018}
R.~Badawy, Y.~P. Raykov, L.~J.~W. Evers, B.~R. Bloem, M.~J. Faber, A.~Zhan,
  K.~Claes, and M.~A. Little, ``{Automated quality control for sensor based
  symptom measurement performed outside the lab},'' \emph{Sensors}, vol.~18,
  no.~4, 2018.

\bibitem{Poorjam2019a}
A.~H. Poorjam, Y.~P. Raykov, R.~Badawy, J.~R. Jensen, M.~G. Christensen, and
  M.~A. Little, ``{Quality control of voice recordings in remote Parkinson's
  disease monitoring using the infinite hidden Markov model},'' in \emph{IEEE
  International Conference on Acoustics, Speech and Signal Processing}, 2019.

\bibitem{Moro-Velazquez2018}
L.~Moro-Vel{\'{a}}zquez, J.~A. G{\'{o}}mez-Garc{\'{i}}a, J.~I. Godino-Llorente,
  J.~Villalba, J.~R. Orozco-Arroyave, and N.~Dehak, ``{Analysis of speaker
  recognition methodologies and the influence of kinetic changes to
  automatically detect Parkinson's disease},'' \emph{Applied Soft Computing},
  vol.~62, pp. 649--666, 2018.

\bibitem{Hermansky1990}
H.~Hermansky, ``{Perceptual linear predictive (PLP) analysis of speech},''
  \emph{The Journal of the Acoustical Society of America}, vol.~87, no.~4, pp.
  1738--1752, 1990.

\bibitem{Orozco-Arroyave2013}
J.~R. Orozco-Arroyave, J.~D. Arias-Londo{\~{n}}o, J.~F. Vargas-Bonilla, and
  E.~N{\"{o}}th, ``{Perceptual analysis of speech signals from people with
  Parkinson's disease},'' \emph{Natural and Artificial Models in Computation
  and Biology - Lecture Notes in Computer Science}, vol. 7930, no.~1, pp.
  201--211, 2013.

\bibitem{Brabenec2017}
L.~Brabenec, J.~Mekyska, Z.~Galaz, and I.~Rektorova, ``{Speech disorders in
  Parkinson's disease: early diagnostics and effects of medication and brain
  stimulation},'' \emph{Journal of Neural Transmission}, vol. 124, no.~3, pp.
  303--334, 2017.

\bibitem{Mekyska2016}
J.~Mekyska, Z.~Smekal, Z.~Galaz, Z.~Mzourek, I.~Rektorova, M.~Faundez-Zanuy,
  and K.~L{\'o}pez-de Ipi{\~n}a, ``Perceptual features as markers of
  {Parkinson's} disease: the issue of clinical interpretability,'' in
  \emph{Recent Advances in Nonlinear Speech Processing}, 2016, pp. 83--91.

\bibitem{Reynolds1995}
D.~Reynolds and R.~Rose, ``{Robust text-independent speaker identification
  using Gaussian mixture speaker models},'' \emph{IEEE Transactions on Speech
  and Audio Processing}, vol.~3, no.~1, pp. 72--83, 1995.

\bibitem{Bot2016}
B.~M. Bot, C.~Suver, E.~C. Neto, M.~Kellen, A.~Klein, C.~Bare, M.~Doerr,
  A.~Pratap, J.~Wilbanks, E.~R. Dorsey, S.~H. Friend, and A.~D. Trister, ``{The
  mPower study, Parkinson disease mobile data collected using ResearchKit},''
  \emph{Scientific Data}, vol.~3, no. 160011, 2016.

\bibitem{vorlander2007auralization}
M.~Vorl{\"a}nder, \emph{Auralization: fundamentals of acoustics, modelling,
  simulation, algorithms and acoustic virtual reality}.\hskip 1em plus 0.5em
  minus 0.4em\relax Springer Science \& Business Media, 2007.

\bibitem{yoshioka2012making}
T.~Yoshioka, A.~Sehr, M.~Delcroix, K.~Kinoshita, R.~Maas, T.~Nakatani, and
  W.~Kellermann, ``Making machines understand us in reverberant rooms:
  Robustness against reverberation for automatic speech recognition,''
  \emph{IEEE Signal Processing Magazine}, vol.~29, no.~6, pp. 114--126, 2012.

\bibitem{Castellano1996}
P.~Castellano, S.~Sradharan, and D.~Cole, ``{Speaker recognition in reverberant
  enclosures},'' in \emph{IEEE International Conference on Acoustics, Speech,
  and Signal Processing Conference Proceedings}, 1996, pp. 117--120.

\bibitem{Allen1979}
J.~B. Allen and D.~A. Berkley, ``{Image method for efficiently simulating
  small-room acoustics},'' \emph{The Journal of the Acoustical Society of
  America}, vol.~65, no.~4, p. 943, 1979.

\bibitem{habets2006room}
E.~A. Habets, ``Room impulse response generator,'' \emph{Technische
  Universiteit Eindhoven, Tech. Rep}, vol.~2, no. 2.4, p.~1, 2006.

\bibitem{eaton2013detection}
J.~Eaton and P.~A. Naylor, ``Detection of clipping in coded speech signals,''
  in \emph{21st European Signal Processing Conference}, 2013, pp. 1--5.

\bibitem{Vincent2007}
E.~Vincent, H.~Sawada, P.~Bofill, S.~Makino, and J.~P. Rosca, ``{First stereo
  audio source separation evaluation campaign: data, algorithms and results},''
  in \emph{Independent Component Analysis and Signal Separation}, M.~E. Davies,
  C.~J. James, S.~A. Abdallah, and M.~D. Plumbley, Eds.\hskip 1em plus 0.5em
  minus 0.4em\relax Berlin, Heidelberg: Springer Berlin Heidelberg, 2007, pp.
  552--559.

\bibitem{hu2007subjective}
Y.~Hu and P.~C. Loizou, ``Subjective comparison and evaluation of speech
  enhancement algorithms,'' \emph{Speech Communication}, vol.~49, no. 7-8, pp.
  588--601, 2007.

\bibitem{Taal2010}
C.~H. Taal, R.~C. Hendriks, R.~Heusdens, and J.~Jensen, ``{A short-time
  objective intelligibility measure for time-frequency weighted noisy
  speech},'' in \emph{IEEE International Conference on Acoustics, Speech and
  Signal Processing}, 2010, pp. 4214--4217.

\bibitem{Habets2007}
E.~Habets, ``{Single- and multi-microphone speech dereverberation using
  spectral enhancement},'' Ph.D. dissertation, Technische Universiteit
  Eindhoven, 2007.

\bibitem{jukic2014speech}
A.~Juki{\'c} and S.~Doclo, ``Speech dereverberation using weighted prediction
  error with laplacian model of the desired signal,'' in \emph{IEEE
  International Conference on Acoustics, Speech and Signal Processing}, 2014,
  pp. 5172--5176.

\bibitem{jukic2015multi}
A.~Juki{\'c}, T.~van Waterschoot, T.~Gerkmann, and S.~Doclo, ``Multi-channel
  linear prediction-based speech dereverberation with sparse priors,''
  \emph{IEEE/ACM Transactions on Audio, Speech and Language Processing},
  vol.~23, no.~9, pp. 1509--1520, 2015.

\bibitem{kameoka2009robust}
H.~Kameoka, T.~Nakatani, and T.~Yoshioka, ``Robust speech dereverberation based
  on non-negativity and sparse nature of speech spectrograms,'' in \emph{IEEE
  International Conference on Acoustics, Speech and Signal Processing}, 2009,
  pp. 45--48.

\bibitem{huang2003class}
Y.~Huang and J.~Benesty, ``A class of frequency-domain adaptive approaches to
  blind multichannel identification,'' \emph{IEEE Transactions on Signal
  Processing}, vol.~51, no.~1, pp. 11--24, 2003.

\bibitem{han2015learning}
K.~Han, Y.~Wang, D.~Wang, W.~S. Woods, I.~Merks, and T.~Zhang, ``Learning
  spectral mapping for speech dereverberation and denoising,'' \emph{IEEE/ACM
  Transactions on Audio, Speech, and Language Processing}, vol.~23, no.~6, pp.
  982--992, 2015.

\bibitem{santos2018speech}
J.~F. Santos and T.~H. Falk, ``Speech dereverberation with context-aware
  recurrent neural networks,'' \emph{IEEE/ACM Transactions on Audio, Speech,
  and Language Processing}, vol.~26, no.~7, pp. 1236--1246, 2018.

\bibitem{srinivasan2006codebook}
S.~Srinivasan, J.~Samuelsson, and W.~B. Kleijn, ``Codebook driven short-term
  predictor parameter estimation for speech enhancement,'' \emph{IEEE
  Transactions on Audio, Speech, and Language Processing}, vol.~14, no.~1, pp.
  163--176, 2006.

\bibitem{he2017multiplicative}
Q.~He, F.~Bao, and C.~Bao, ``Multiplicative update of auto-regressive gains for
  codebook-based speech enhancement,'' \emph{IEEE/ACM Transactions on Audio,
  Speech, and Language Processing}, vol.~25, no.~3, pp. 457--468, 2017.

\bibitem{mohammadiha2013supervised}
N.~Mohammadiha, P.~Smaragdis, and A.~Leijon, ``Supervised and unsupervised
  speech enhancement using nonnegative matrix factorization,'' \emph{IEEE
  Transactions on Audio, Speech, and Language Processing}, vol.~21, no.~10, pp.
  2140--2151, 2013.

\bibitem{wang2018supervised}
D.~Wang and J.~Chen, ``Supervised speech separation based on deep learning: An
  overview,'' \emph{IEEE/ACM Transactions on Audio, Speech, and Language
  Processing}, vol.~26, no.~10, pp. 1702--1726, 2018.

\bibitem{kavalekalam2019model}
M.~S. Kavalekalam, J.~K. Nielsen, J.~B. Boldt, and M.~G. Christensen,
  ``Model-based speech enhancement for intelligibility improvement in binaural
  hearing aids,'' \emph{IEEE/ACM Transactions on Audio, Speech and Language
  Processing}, vol.~27, no.~1, pp. 99--113, 2019.

\bibitem{goh1999kalman}
Z.~Goh, K.~C. Tan, and B.~T.~G. Tan, ``Kalman-filtering speech enhancement
  method based on a voiced-unvoiced speech model,'' \emph{IEEE Transactions on
  Speech and Audio Processing}, vol.~7, no.~5, pp. 510--524, 1999.

\bibitem{nielsen2017fast}
J.~K. Nielsen, T.~L. Jensen, J.~R. Jensen, M.~G. Christensen, and S.~H. Jensen,
  ``Fast fundamental frequency estimation: Making a statistically efficient
  estimator computationally efficient,'' \emph{Signal Processing}, vol. 135,
  pp. 188--197, 2017.

\bibitem{varga1993assessment}
A.~Varga and H.~J. Steeneken, ``Assessment for automatic speech recognition:
  {II. NOISEX-92}: A database and an experiment to study the effect of additive
  noise on speech recognition systems,'' \emph{Speech Communication}, vol.~12,
  no.~3, pp. 247--251, 1993.

\bibitem{erkelens2007minimum}
J.~S. Erkelens, R.~C. Hendriks, R.~Heusdens, and J.~Jensen, ``Minimum
  mean-square error estimation of discrete {Fourier} coefficients with
  generalized {Gamma} priors,'' \emph{IEEE Transactions on Audio, Speech, and
  Language Processing}, vol.~15, no.~6, pp. 1741--1752, 2007.

\bibitem{gerkmann2012unbiased}
T.~Gerkmann and R.~C. Hendriks, ``Unbiased {MMSE}-based noise power estimation
  with low complexity and low tracking delay,'' \emph{IEEE Transactions on
  Audio, Speech, and Language Processing}, vol.~20, no.~4, pp. 1383--1393,
  2012.

\bibitem{habets2008joint}
E.~A. Habets, S.~Gannot, I.~Cohen, and P.~C. Sommen, ``Joint dereverberation
  and residual echo suppression of speech signals in noisy environments,''
  \emph{IEEE Transactions on Audio, Speech, and Language Processing}, vol.~16,
  no.~8, pp. 1433--1451, 2008.

\bibitem{kodrasi2016joint}
I.~Kodrasi and S.~Doclo, ``Joint dereverberation and noise reduction based on
  acoustic multi-channel equalization,'' \emph{IEEE/ACM Transactions on Audio,
  Speech and Language Processing}, vol.~24, no.~4, pp. 680--693, 2016.

\bibitem{Reynolds2000}
D.~A. Reynolds, T.~F. Quatieri, and R.~B. Dunn, ``{Speaker verification using
  adapted Gaussian mixture models},'' \emph{Digital Signal Processing},
  vol.~10, no.~1, pp. 19--41, 2000.

\bibitem{Deller2000}
J.~R. Deller, J.~H.~L. Hansen, and J.~G. Proakis, \emph{{Discrete-time
  processing of speech signals}}, 2nd~ed.\hskip 1em plus 0.5em minus
  0.4em\relax New York: IEEE Press, 2000.

\bibitem{Poorjam2018}
A.~H. Poorjam, M.~A. Little, J.~R. Jensen, and M.~G. Christensen, ``{A
  supervised approach to global signal-to-noise ratio estimation for whispered
  and pathological voices},'' in \emph{IEEE International Conference on
  Acoustics, Speech and Signal Processing}, 2018.

\bibitem{Jeub2009}
M.~Jeub, M.~Schafer, and P.~Vary, ``{A binaural room impulse response database
  for the evaluation of dereverberation algorithms},'' in \emph{International
  Conference on Digital Signal Processing}, 2009, pp. 1--5.

\bibitem{Schroeder1985}
M.~Schroeder and B.~Atal, ``{Code-excited linear prediction(CELP): High-quality
  speech at very low bit rates},'' \emph{IEEE International Conference on
  Acoustics, Speech, and Signal Processing}, vol.~10, pp. 937--940, 1985.

\bibitem{Rabiner1989}
L.~R. Rabiner, ``{A tutorial on hidden Markov models and selected applications
  in speech recognition},'' \emph{Proceedings of the IEEE}, vol.~77, no.~2, pp.
  257--286, 1989.

\bibitem{Sethuraman1994}
J.~Sethuraman, ``{A constructive definition of Dirichlet priors},''
  \emph{Statistica Sinica}, vol.~4, pp. 639--650, 1994.

\end{thebibliography}
